\documentclass[10pt,preprint]{aastex}
\usepackage{emulateapj5,apjfonts,onecolfloat5}

\newcommand{\axp}{1E\, 1841-045}
\newcommand{\axprxsfl}{1RXS\, J170849.0\,-\,400910}
\newcommand{\axprxs}{1RXS\, J1708\,-\,4009}
\newcommand{\axpu}{4U\, 0142\,$+$\,61}
\newcommand{\axpe}{1E\, 2259\,$+$\,586}
\newcommand{\axpee}{1E\, 1048.1\,-\,5937}

\newcommand{\gr}{$\gamma$-ray}

\newcommand{\gtap}{\mathrel{\hbox{\rlap{\lower.55ex \hbox {$\sim$}}
                   \kern-.3em \raise.4ex \hbox{$>$}}}}
\newcommand{\ltap}{\mathrel{\hbox{\rlap{\lower.55ex \hbox {$\sim$}}
                   \kern-.3em \raise.4ex \hbox{$<$}}}}

\slugcomment{}

\shorttitle{Detection of hard soft \gr\ emission from AXPs}
\shortauthors{Kuiper et al.}

\begin{document}

\twocolumn[

\doublespace


\title{Discovery of luminous pulsed hard X-ray emission from 
anomalous X-ray pulsars \axprxs, \axpu\ and \axpe\ by INTEGRAL and RXTE}

\author{L. Kuiper\altaffilmark{1}, W. Hermsen\altaffilmark{1,2}, P.R. den 
Hartog\altaffilmark{1} and W. Collmar\altaffilmark{3}}
\email{L.M.Kuiper@sron.nl}
\altaffiltext{1}{SRON-National Institute for Space Research, Sorbonnelaan 2, 
3584 CA, Utrecht, The Netherlands }
\altaffiltext{2}{Astronomical Institute ``Anton Pannekoek", University of 
Amsterdam, Kruislaan 403, 1098 SJ Amsterdam, The Netherlands}
\altaffiltext{3}{Max-Planck-Institut f\"ur extraterrestrische Physik, P.O. Box 
1312, 85741 Garching, Germany}


\begin{abstract}
Triggered by the earlier surprising detection of pulsed hard X-ray emission 
from \axp, we investigated the time-averaged high-energy spectral characteristics 
of the established Anomalous X-ray Pulsars, \axprxs, \axpu, \axpe\ and \axpee, 
all with persistent X-ray emission.  We report on the discovery of hard spectral 
tails for energies above 10 keV in the total and pulsed spectra of AXPs \axprxs,
\axpu\ and \axpe\ using RXTE PCA (2-60 keV) / HEXTE (15-250 keV) data and INTEGRAL 
IBIS ISGRI (20-300 keV) data. \axpee\ appeared to be too weak to be detected with
the presently available exposure. Furthermore, improved spectral information for 
\axp\ is presented. The pulsed and total spectra measured above 10 keV have power-law shapes and
there is so far no significant evidence for spectral breaks or bends up to $\sim$ 150 keV.
The pulsed spectra above 10 keV are exceptionally hard with indices measured for 
four AXPs approximately in the range -1.0 -- 1.0. The indices measured for these pulsed spectra below 10 keV
were in the range 2.0 -- 4.3, indicating the very drastic spectral changes in a narrow
energy interval around 10 keV, where clear minima are found in luminosity. The best 
fit power-law models to the total spectra between $\sim$ 10 and 150 keV are significantly
softer, with indices measured for three AXPs, \axp, \axprxs\ and \axpu\, in the range 1.0 -- 1.4.
For the latter AXPs the pulsed fractions are consistent with 100\% around 100 keV, but
are different at 10 keV: \axpu\ $\sim$ 10\%, \axp\ $\sim$25\%, and \axprxs\ consistent
with 100\%. The luminosities of these total and pulsed spectral tails (10- 150 keV) largely 
exceed the total available spin-down powers by factors ranging from $\sim 100$ to $\sim 600$.
This shows that also these new hard-X-ray components cannot be powered by rotational energy loss.
We reanalyzed archival CGRO COMPTEL (0.75-30 MeV) data to search for signatures from our set of AXPs. 
No detections can be claimed, but the obtained upper-limits to soft gamma-ray emission in this MeV range 
indicates for \axprxs, \axpu\ and \axp\ that strong breaks or bends must occur somewhere between 
$\sim$ 150 keV and 750 keV. 
We discuss predictions from first attempts to model our hard X-ray / soft gamma-ray spectra 
in the context of the magnetar model. Our spectral results cannot yet discriminate between the 
different proposed scenarios. Particularly, more constraining information is required on the 
detailed shape of the spectra between $\sim$ 150 keV (our highest data points from INTEGRAL) and 
750 keV, from where we report the COMPTEL upper limits.  
\end{abstract}

\keywords{pulsars: individual (\axprxs; \axpu; \axpe;\axpee;\axp), X-rays: stars}

]


\section{Introduction}

Anomalous X-ray pulsars (AXP) belong to a class of rare objects closely concentrated along the 
Galactic plane, known to emit pulsed X-rays for energies below 10 keV with pulse periods in the range $\sim$ 6 - 
12 s and characteristic spin-down time scales of $\sim 10^3$ - $10^5$ year. Two, probably three, members 
of this class are embedded in a shell-like supernova remnant \citep[SNR; see e.g.][]{gregory80,kriss85}.
The fact that the observed X-ray luminosity is much larger than the spin-down 
power excludes an interpretation in which the (pulsed) X-ray emission originates from a spin-down 
powered pulsar. On the other hand, the steady spin-down without a doppler modulated signature 
and the lack of bright optical counterparts make an X-ray binary interpretation, in which mass 
transfer (accretion) powers the high-energy emission of the system, very unlikely \citep[see][for a review 
on AXPs]{mereghetti02}. Currently, models based on the decay of very strong magnetic fields 
($10^{14}$-$10^{15}$ Gau\ss) - so called ``magnetar'' models \citep{thompson96} - seem to explain the 
observed high-energy characteristics of AXPs at a satisfactory level. For instance, the recently detected bursts 
from the AXPs \axpee\ \citep{gavriil02b} and \axpe\ \citep{kaspi03} mimic the bursting behaviour of 
Soft Gamma-ray Repeaters (SGR) for which the magnetar model was initially developed. 
Also, the ``glitch'' phenomenon detected in the spin-down of some AXP members fits in this model \citep{kaspi00,kaspi03,morii05}. 
These properties provide strong evidence that both AXPs and SGRs are members of the same source class.

The X-ray spectra of AXPs in the 0.5-10 keV band are very soft and can best be described by a black body plus a 
power-law model. The softness of the spectra below 10 keV (power-law indices $2 < \Gamma < 4$, with $F_{\gamma} 
\propto E^{-\Gamma}$) predicts non-detections for energies above 10 keV and thus explains the initial ignorance for 
studying the spectral properties of AXPs at energies above 10 keV.

It was a great surprise that the high-resolution imaging instrument IBIS/ISGRI aboard ESA's INTEGRAL satellite 
measured hard X-rays from the direction of three AXPs. Firstly, \citet{molkov04} reported 
the discovery of an INTEGRAL source at the position of AXP \axp\ in SNR Kes 73 for energies up to 120 keV
(60--120 keV: 7.5 $\sigma$). This was followed up by \citet{kuiper04}, who analysed 
archival RXTE PCA and HEXTE data from monitoring observations spread over four years, to 
prove that the hard X-ray emission comes from the AXP and not from the SNR. They discovered 
non-thermal pulsed hard X-ray / soft $\gamma$-ray emission up to $\sim$ 150 keV with a spectrum with 
power-law photon index of $\sim$ 0.94. 

Secondly, \citet{revnivtsev04} published the INTEGRAL detection (18--60 keV: 6.5 $\sigma$) of AXP \axprxsfl\ 
(we use throughout this paper: \axprxs), also in a spatial analysis of ISGRI data. Similarly, 
\citet{hartog04} reported the detection of AXP \axpu\ (50--100 keV: 6.1 $\sigma$) which exhibited a very hard 
total spectrum above 20 keV. These three AXPs reach about the same flux level around 100 keV.
 
In this paper we present the results from follow-up studies using archival RXTE PCA and 
HEXTE data of \axprxs, \axpu, \axpe\ and \axpee\ aimed at studying their timing and spectral characteristics 
above 10 keV. We also revisit the pulsed high-energy emission properties of \axp\ above 10 keV using more
RXTE PCA data and applying now more-optimized event selection criteria. Furthermore, we explored the INTEGRAL 
database using both public, private and core program data to derive the total X-ray emission spectra of the
five AXPs mentioned above, and reanalyze archival data from COMPTEL \citep{schonfelder93}, delivering constraining 
upper limits in the soft gamma-ray band 0.75-30 MeV. Finally, initial results will be shown from IBIS ISGRI timing 
analysis studies of \axprxs\ and \axp.   

\begin{table}[t]
\caption{List of RXTE observations of \axprxs, \axpu, \axpe\ and \axpee\ used in 
this study. For observations used in the reanalysis of \axp\ see Sect. \ref{sect_pca_tm} and \citet{kuiper04}.\label{obs_table}}
{\footnotesize
\begin{center}
\begin{tabular}{cccr}
\hline
\textbf{Obs.} & \multicolumn{2}{c}{\textbf{Begin/End Date}}    & 
\textbf{Exp.$^{\dagger}$}\\
\textbf{id.}  & \multicolumn{2}{c}{\textbf{(dd/mm/yyyy)}}      & \textbf{(ks)}\\
\hline
\multicolumn{4}{l}{\textit{\axprxs}}\\
30125            & 12-01-1998          & 08-01-1999           &  59.896 \\
40083            & 06-02-1999          & 11-03-2000           &  52.568 \\
50082            & 21-04-2000          & 12-05-2001           &  34.576 \\
60412            & 20-05-2001          & 23-05-2001           &   9.928 \\
60069            & 06-05-2001          & 20-02-2002           &  24.544 \\
70094            & 02-04-2002          & 20-03-2003           &  55.600 \\
80098            & 16-04-2003          & 26-10-2003           &  73.368 \\
All              & 12-01-1998          & 26-10-2003           & 310.480 \\
\hline\hline
\multicolumn{4}{l}{\textit{\axpu}}\\
10193            & 28-03-1996          & 29-03-1996           &  37.728 \\
10185            & 28-03-1996          & 28-03-1996           &  16.288 \\
20146            & 24-11-1996          & 13-12-1997           &  10.600 \\
30110            & 21-03-1998          & 21-03-1998           &  15.408 \\
50082            & 07-03-2000          & 10-02-2001           &  34.240 \\
60069            & 18-03-2001          & 08-01-2002           &  42.272 \\
70094            & 06-03-2002          & 26-12-2002           &  82.464 \\
80098            & 28-03-2003          & 18-09-2003           &  38.664 \\
80099            & 03-09-2003          & 09-09-2003           &  29.216 \\
All              & 28-03-1996          & 18-09-2003           & 306.880 \\
\hline\hline
\multicolumn{4}{l}{\textit{\axpe}}\\
10192            & 29-09-1996          & 30-09-1996           &  75.936 \\ 
20145            & 25-02-1997          & 25-03-1997           & 101.024 \\ 
20146            & 24-11-1996          & 13-12-1997           &   9.784 \\ 
30126            & 13-08-1998          & 02-12-1998           & 103.136 \\ 
40083            & 17-01-1999          & 01-03-2001           &  48.344 \\ 
40082            & 26-01-2000          & 27-03-2000           &  87.120 \\ 
50082            & 10-03-2000          & 02-03-2001           &  50.280 \\ 
60069            & 16-04-2001          & 07-02-2002           &  49.592 \\ 
70094            & 22-03-2002          & 15-02-2003           & 149.840 \\ 
80098            & 15-03-2003          & 28-10-2003           &  71.960 \\ 
All              & 29-09-1996          & 28-10-2003           & 747.016 \\ 
\hline\hline
\multicolumn{4}{l}{\textit{\axpee}}\\
10192            & 29-07-1996          & 30-07-1996           &  72.344 \\ 
20146            & 24-11-1996          & 13-12-1997           &  12.616 \\ 
40083            & 23-01-1999          & 10-02-2000           &  49.184 \\ 
50082            & 11-03-2000          & 09-02-2001           &  40.992 \\ 
60069            & 06-03-2001          & 25-02-2002           & 139.776 \\ 
70094            & 12-03-2002          & 25-02-2003           & 133.224 \\ 
80098            & 12-03-2003          & 24-02-2004           & 136.376 \\ 
All              & 29-07-1996          & 24-02-2004           & 584.512  \\ 
\hline
\multicolumn{4}{l}{$^{\dagger}$PCU-2 exposure after screening} \\
\end{tabular}
\end{center}}
\end{table}

\section{Instruments and observations}

\subsection{Rossi X-ray Timing Explorer}

In this study extensive use is made of data from monitoring observations
of AXPs with the two non-imaging X-ray instruments aboard RXTE, the Proportional Counter Array 
(PCA; 2-60 keV) and the High Energy X-ray Timing Experiment (HEXTE; 15-250 keV). The PCA 
\citep{jahoda96} consists of five collimated xenon proportional 
counter units (PCUs) with a total effective area of $\sim 6500$ cm$^2$ over a $\sim 1\degr$ 
(FWHM) field of view. Each PCU has a front Propane anti-coincidence layer and three Xenon 
layers which provide the basic scientific data, and is sensitive to 
photons with energies in the range 2-60 keV. The energy resolution is about 18\% at 6 keV.

The HEXTE instrument \citep{rothschild98} consists of two independent detector 
clusters, each containing four Na(Tl)/ CsI(Na) scintillation
detectors. The HEXTE detectors are mechanically collimated to a $\sim 1\degr$ (FWHM) 
field of view and cover the 15-250 keV energy range with an energy resolution of 
$\sim$ 15\% at 60 keV. The collecting area is 1400 cm$^2$ taking into account the 
loss of the spectral capabilities of one of the detectors. The maximum time 
resolution of the tagged events is $7.6\mu$s. In its default operation mode the 
field of view of each cluster is switched on and off source to provide instantaneous 
background measurements.
Due to the co-alignment of HEXTE and the PCA, they simultaneously observe the 
sources. Table \ref{obs_table} lists the publicly available RXTE 
observations used in this study. In the fourth column the PCU unit-2 screened 
(see Sect \ref{rxte_timing}) exposure is given. A typical observation consists of 
several sub-observations spaced more or less uniformly between the start and end date of 
the observation.

\subsection{INTEGRAL}

The INTEGRAL spacecraft \citep{winkler03}, launched 17 October 2002, carries two main 
$\gamma$-ray instruments: a high-angular-resolution imager IBIS \citep{ubertini03} and
a high-energy-resolution spectrometer SPI \citep{vedrenne03}. These instruments make use of 
coded aperture masks enabling image reconstruction in the hard X-ray/soft $\gamma$-ray band.

In our study, guided by sensitivity considerations, we only used data recorded by the INTEGRAL 
Soft Gamma-Ray Imager ISGRI \citep{lebrun03}, the upper detector system of IBIS, sensitive 
to photons with energies in the range $\sim$20 keV -- 1 MeV. With an angular resolution of about $12\arcmin$
and a source location accuracy of better than $1\arcmin$ (for a $>10\sigma$ source) this instrument
is able to locate and separate high-energy sources in crowded fields within its $19\degr \times 19\degr$ 
field of view (50\% partially coded) with an unprecedented sensitivity ($\sim$ 960 cm$^2$ at 50 keV).
Its energy resolution of about 7\% at 100 keV is amply sufficient to determine the (continuum) spectral 
properties of hard X-ray sources in the $\sim$ 20 - 300 keV energy band.

The timing accuracy of the ISGRI time stamps recorded on board is about $61\mu$s. The time 
alignment between INTEGRAL and RXTE is better than $\sim 25\mu$s, verified using data 
from simultaneous RXTE and INTEGRAL observations of the accretion-powered millisecond pulsar 
IGR J00291+5934 \citep{falanga05}; for a calibration on the Crab pulsar, see also \citet{kuiper03}. 
Given the fact that the accuracy of the RXTE clock in absolute time is about $2\mu$s \citep{rots04},
this implies that the INTEGRAL absolute timing is better than $\sim 27 \mu$s. 
Data from regular INTEGRAL Crab monitoring observations show that the clock behaviour is stable, making 
deep timing studies of weak pulsars possible.

In its default operation mode INTEGRAL observes the sky in a dither pattern 
with $2\degr$ steps, which could be rectangular e.g. a $5 \times 5$ dither pattern 
with 25 grid points, or hexagonal with 7 grid points (target in the middle). Typical integration
times for each grid point (pointing/sub-observation) are in the range 1800 - 
3600 seconds. This strategy drives the structure of the INTEGRAL data archive which is 
organised in so-called science windows (Scw) per INTEGRAL orbital revolution (lasting for 
about 3 days) containing the data from all instruments for a given pointing. 
Most of the INTEGRAL data reduction in this study was performed with the Offline 
Scientific Analysis (OSA) version 4.1 distributed by the INTEGRAL Science Data Centre 
\citep[ISDC; see e.g.][]{courvoisier03}.

Table \ref{obsint_table} lists the INTEGRAL orbital revolution identifiers with corresponding start/end dates of the
observations used in the imaging/spectral analyses and timing analyses of the selected sample of persistent AXPs.

\begin{table}[t]
\caption{List of INTEGRAL observations, sorted on INTEGRAL orbital revolutions (Rev.), 
of the AXPs studied in this work. For more details on the executed INTEGRAL observations, see 
{http://integral.esac.esa.int/} \label{obsint_table}}
{\footnotesize
\begin{center}
\begin{tabular}{llcc}
\hline
\textbf{Rev.}   & \textbf{Rev.}   & \multicolumn{2}{c}{\textbf{Begin/End Date}}    \\
\textbf{begin}  & \textbf{end}    & \multicolumn{2}{c}{\textbf{(dd/mm/yyyy)}}      \\
\hline\hline
\multicolumn{4}{c}{\textbf{Imaging analysis}}\\
\hline\hline
\multicolumn{4}{l}{\textit{\axprxs}}\\
 36           & 106          & 29-01-2003           & 29-08-2003  \\
\hline
\multicolumn{4}{l}{\textit{\axpu}}\\
 47           &  92          & 03-03-2003           & 16-07-2003  \\
142           & 148          & 12-12-2003           & 01-01-2004  \\
153           & 153          & 14-01-2004           & 15-01-2004  \\
161           & 162          & 07-02-2004           & 12-02-2004  \\
177           & 234          & 26-03-2004           & 12-09-2004  \\
238           & 238          & 25-09-2004           & 27-09-2004  \\
261           & 266          & 02-12-2004           & 19-12-2004  \\
268           & 269          & 24-12-2004           & 28-12-2004  \\
\hline
\multicolumn{4}{l}{\textit{\axp}}\\
 49           & 253          & 10-03-2003           & 08-11-2004  \\
\hline
\multicolumn{4}{l}{\textit{\axpe}}\\
142           & 148          & 12-12-2003           & 01-01-2004  \\
161           & 162          & 07-02-2004           & 12-02-2004  \\
\hline
\multicolumn{4}{l}{\textit{\axpee}}\\
 36           & 217          & 29-01-2003           & 24-07-2004  \\
\hline\hline
\multicolumn{4}{c}{\textbf{Timing analysis}}\\
\hline\hline
\multicolumn{4}{l}{\textit{\axprxs}}\\
 36           & 120          & 29-01-2003           & 10-10-2003  \\
\hline
\multicolumn{4}{l}{\textit{\axp}}\\
 49           & 123          & 10-03-2003           & 18-10-2003  \\
\hline
\end{tabular}
\end{center}}
\end{table}

 \subsection{COMPTEL}  
 
The Compton telescope COMPTEL aboard the Compton Gamma-Ray Observatory (CGRO, 1991 -- 2000) 
was sensitive to $\gamma$-ray photons between 0.75 and 30 MeV, thereby covering the harder 
$\gamma$-ray band adjacent to the INTEGRAL one. The very hard spectra that we measured with 
IBIS ISGRI for some AXPs warranted us to revisit the COMPTEL data archive to search for 
signals from AXPs. COMPTEL has an energy-dependent energy and angular resolution of 5\% -- 8\% 
(FWHM) and $1\fdg7$ -- $4\fdg4$ (FWHM), respectively, and a wide circular field of view 
covering $\sim$1 steradian. Imaging in its large field of view is possible with 
a location accuracy (flux dependent) of the order of $0\fdg5$ -- $2\degr$. 
For details on the experiment see \citet{schonfelder93}. 

\section{Analysis methods}
  \subsection{RXTE PCA/HEXTE timing}
  \label{rxte_timing}
  The PCA data from the observations listed in Table \ref{obs_table} have all been collected in {\em Goodxenon} or 
  {\em GoodxenonwithPropane} mode allowing high-time-resolution ($0.9\mu$s) studies in 256 spectral channels. 
  Because we are mainly interested in the medium/hard ($>2$ keV) X-ray timing properties of the selected sample of AXPs, 
  we ignored the events triggered in the Propane layers of each PCU. 
  Furthermore, contrary to the work presented in \citet{kuiper04}, we now used data from {\em all} three xenon layers of each 
  PCU, namely, employing data from the (deeper) middle and lower xenon layers considerably improves the signal-to-noise 
  ratio for energies above $\sim 10$ keV. This allows us to better characterize the hard ($>10$ keV) X-ray properties of AXPs. 

  The number of active PCUs at any time was changing, therefore, we treated the five PCUs constituting the PCA separately. 
  Good time intervals have been determined for each PCU by including only time periods when the PCU in question is on, 
  and during which the pointing direction is within $0\fdg 05$ from the target, the elevation angle above Earth's horizon 
  is greater than $5\degr$, a time delay of 30 minutes since the peak of a South-Atlantic-Anomaly passage holds, and a 
  low background level due to contaminating electrons is observed. 
  These good time intervals have subsequently been applied in the screening process to the data streams from each of 
  the PCUs (e.g. see Table \ref{obs_table} for the resulting screened exposure of PCU-2 per observation run).

  Next, for each sub-observation the arrival times of the selected events (for each PCU unit) have been converted to 
  arrival times at the solar system barycenter (in TDB (=barycentric dynamical time) time scale; DE200 solar system ephemeris) 
  using the instantaneous spacecraft position and the celestial positions of the selected sample of AXPs. In this work we used 
  for the AXP positions those obtained by the Chandra X-ray observatory with a typical position accuracy of about $0\farcs5$ (for \axprxs, 
  see \citet{israel03}; for \axpu, see \citet{patel03}; for \axpe, see \citet{patel01}; for \axpee, see \citet{wang02,israel02} 
  ; and finally for \axp, see \citet{wachter04}).

  These barycentered arrival times have been folded with available phase connected timing solutions (see for details the 
  relevant section on the timing characteristics for each AXP) using only the first three frequency coefficients (frequency, 
  first and second frequency time derivatives at a certain epoch) to obtain pulse phase distributions for selected energy windows.  
  Combining now the phase distributions from the various PCUs for energies between $\sim 2$ and 10 keV we obtained for each 
  sub-observation pulse profiles, which deviate from uniformity at significances well above $5\sigma$ for each of the AXPs in our sample.
  For the calculation of these significances we applied in this work the {\em bin-free} $Z_{n}^2$ statistic \citep{buccheri83}, which
  behaves as a $\chi^2$ distribution for $2n$ degrees of freedom (n=number of harmonics).
  However, phase shifts between the sub-observations made a direct combination of the pulse profiles to obtain (very) 
  high-statistics {\em time-averaged} pulse profiles impossible. Therefore we correlated the pulse phase distribution of each  
  sub-observation with a chosen initial template and applied the measured phase shifts to obtain an aligned combination with much 
  higher statistics. The correlation is then once repeated with, instead of the initially chosen template, the aligned
  combination from the first correlation to obtain the final summed profile \citep[see e.g.][for a similar iterative method applied 
  for PSR B0540-69]{deplaa03}.

  The net result is a set of aligned high-statistics pulse profiles in 256 energy channels for each of the AXPs in our sample. 
  It should be noted that these profiles are {\em time-averaged} profiles ignoring possible temporal variations in shape and/or 
  pulsed fraction (see e.g. \citet{rea05} for \axprxs; \citet{rea06} for \axpu; and \citet{gavriil04b,tiengo05} for \axpee).

  HEXTE operated in its default rocking mode during the observations listed in Table \ref{obs_table}, allowing the 
  collection of real-time background data from two independent positions $\pm 1\fdg 5$ to either side of the 
  on-source position. For the timing analysis we selected only the on-source data.
  Good time intervals have been determined using similar screening filters as used in the case of the PCA. The 
  selected on-source HEXTE event times have subsequently been barycentered and folded according to the available ephemerides 
  (see PCA part at the beginning of this section) using again only the first three frequency coefficients.
  Applying, for each AXP in our sample, the phase shifts as derived from the contemporaneous PCA measurements to the HEXTE phase 
  distributions of each sub-observation we obtained the time averaged HEXTE pulse phase distributions in 256 spectral channels 
  (15 - 250 keV) for the combination of observations listed in Table \ref{obs_table}.   

  \subsection{RXTE PCA/HEXTE spectral analysis}
  \label{rxte_spectral}

  Because of the non-imaging nature of the two main RXTE instruments the (total) source-flux estimation relies on accurate 
  time-dependent instrumental and celestial background measurements. Although such models exist for the PCA, the complexity of
  the near environment of the AXPs in our sample makes it very difficult to derive reliable unbiased total flux estimates with 
  the PCA in the 2-30 keV range. Specifically, all AXPs are located in a narrow strip along the Galactic plane where (large)
  gradients in the Galactic ridge emission exist \citep{valinia98}; \axp\ and \axpe\ are located in supernova remnants; 
  time-variable strong sources are present near to \axpu\ (Be X-ray binary RX J0146.9+6121), \axpee\ (the enigmatic 
  $\eta$ Carina) and \axprxs\ (e.g. the strong and highly variable X-ray binaries OAO 1657-415 and 4U 1700-377). 

  The rocking strategy applied during HEXTE operations in principle provides instantaneous background measurements, but also in this
  case the gradient in the Galactic ridge emission and the possible presence of other (strong) sources in both the on and off-source 
  pointings (very serious for e.g. \axprxs\ and \axpu) can result in unreliable (background subtracted) total-source-flux measurements.
  Therefore we abandoned, in contrast to the work presented in \citet{kuiper04} for \axp\ in Kes 73, the derivation of the total-source 
  flux with the non-imaging PCA and HEXTE instruments.

  In this work we concentrate on the derivation of the {\em time-averaged pulsed\/} PCA/HEXTE spectra of the AXPs in our sample. 
  This can be done by determining the number of pulsed counts in differential PCA/HEXTE energy bands by fitting a truncated Fourier 
  series \begin{equation} N(\phi) = a_0+\sum_{k=1}^{N} a_k \cos(2\pi k \phi) + b_k \sin(2\pi k \phi) \label{eq_1} \end{equation} 
  with $\phi$ the pulse phase, to the measured pulse phase distributions $N(\phi)$. It turned out that 3 to 5 harmonics ($N=3/5$) 
  were sufficient to describe the measured distributions accurately for all energy intervals and AXPs in our sample. In the case of 
  the PCA we derived for each PCU the energy response matrix (energy redistribution including the sensitive area) for the combination 
  of observations listed in Table \ref{obs_table} and subsequently took the different PCU (screened) exposure times into account in the 
  construction of the weighted PCU-combined energy response. 
  The pulsed (excess) counts per energy band are fitted in a procedure assuming either an absorbed power-law
  ($F_{\gamma}=K\cdot e^{-N_H\cdot \sigma} \cdot E_{\gamma}^{-\Gamma}$), or an absorbed double power-law
  ($F_{\gamma}=e^{-N_H \cdot \sigma}\cdot (K_1 \cdot E_{\gamma}^{-\Gamma_1} + K_2 \cdot E_{\gamma}^{-\Gamma_2})$) or an absorbed black body plus
  power-law ($F_{\gamma}=e^{-N_H\cdot \sigma}\cdot (K \cdot E_{\gamma}^{-\Gamma}+ K_{bb} \cdot E_{\gamma}^2 / (\exp(E_{\gamma}/kT)-1))$) photon spectrum folded through the PCU-combined energy response. 

  In the spectral ``deconvolution" process of the HEXTE total pulsed counts in almost all cases\footnote{For HEXTE pointings with the
  target (AXP) far off-axis e.g. \axpu\ during an observation of HMXB RX J0146.9+6121, we took the reduction in the effective sensitive area due 
  to the collimator response into account} the on-axis cluster A and B energy response matrices have been employed taking into account 
  the (slightly) different screened on-source exposure times for each cluster. 
  The exposure times have been corrected for the considerable deadtime effects.

  \subsection{INTEGRAL timing analysis}
  \label{sect_int_timing}
  The first step in an INTEGRAL timing analysis is to obtain a set of science windows for which the angular distance between instrument 
  pointing direction and target is within $14\fdg5$ to ensure that (a part of) the detector plane is illuminated by the target. The resulting
  list is further screened on erratic (ISGRI) count rate variations, indicative for particle effects due to Earth radiation belt 
  passages or solar flare activities. These science windows are excluded for further analysis.
  Next, only events with rise times between 7 and 90 \citep[see][for definition]{lebrun03}, detected in {\em non-noisy} 
  ISGRI detector pixels which have an illumination factor of more than 25\% (i.e. at least 25\% of a detector pixel must have been illuminated 
  by the target) are passed for further analysis.
  The on-board event time stamps are corrected for known instrumental (fixed), ground station and general time delays in the on-board time vs.
  TT (Terrestrial Time) correlation \citep[see e.g.][]{walter03}.
  The resulting event times in TT of the selected events are barycentered (using the JPL DE200 solar system ephemeris) adopting the Chandra X-ray 
  positions of the AXPs and the instantaneous INTEGRAL orbit information.
  These barycentered events are finally folded using an appropriate timing model ($\nu,\dot\nu,\ddot\nu$ and the epoch) 
  to yield pulse phase distributions for different energy bands between 20 and 300 keV.
  The timing models (phase connecting ephemerides) are based on publicly available RXTE monitoring data of AXPs.
  Ephemerides have been generated for two AXPs in our sample, \axprxs\ and \axp, because at the INTEGRAL epoch (MJD $> 52668$ and MJD $> 52698$ for 
  \axprxs\ and \axp, respectively) these were not available from existing literature (see Table \ref{eph_table}; all with very small RMS values, 
  required for extracting the weak pulsed signals). 
 
\begin{table*}[t]
\caption{Phase coherent ephemerides for \axprxs\ and \axp, derived from RXTE PCA monitoring data and valid for the 
analyzed INTEGRAL observations.\label{eph_table}}
{\footnotesize
\begin{center}
\begin{tabular}{lccccccc}
\hline
AXP & Start &  End  & Epoch   & $\nu$  & $\dot\nu$              & $\ddot\nu$     
            & RMS\\
    & [MJD] & [MJD] & [MJD,TDB]   & [Hz]   & $\times 10^{-13}$ Hz/s & $\times 
10^{-22}$ Hz/s$^2$ &    \\
\hline
\hline
\axprxs & 52590          & 52939        & 52590.0          & 0.09089812328(36) & -1.59836(30) & 0.00(Fixed)  & 0.012\\
\axp    & 52726          & 52982        & 52726.0          & 0.0848972590(50)  & -3.2217(85)  & 4.22(83) & 0.015\\
\hline
\hline
\end{tabular}
\end{center}}
\end{table*}

\subsection{INTEGRAL spectral analysis}

The INTEGRAL IBIS ISGRI spectral analysis applied in our study is based on OSA 4.1 programs producing sky mosaics for the 
combination of many science windows in different energy bands \citep{goldwurm03}. The resulting dead-time corrected ISGRI 
source-count rates per energy band are referenced to count rates measured from the Crab in similar energy bands (Crab 
calibration observations during INTEGRAL Revs. 102/103 were used). 

We used for the total Crab photon emission spectrum the following spectrum derived by \citet{willingale01} based on XMM-Newton observations 
of the Crab at energies between 0.3 and 10 keV: photon index $\Gamma=2.108(6)$ with a normalization at 1 keV of $9.59(5)$
ph/(cm$^2$s keV). 

We verified the validity of the extrapolation of this spectrum to energies between 15 and 250 keV using RXTE HEXTE Crab data. 
The HEXTE data utilized is from a long dedicated RXTE Crab observation (obs. id. 40805), performed between
17-31 March 1999 and 18-19 Dec. 1999, yielding dead time corrected cluster 0 and 1 exposures of 22.7 and 23.8 ks, respectively.
Applying an overall (energy independent) normalization factor of 1.087 to be consistent with the BeppoSAX LECS/MECS/PDS spectrum \citep{kuiper01}
the derived 15-250 keV HEXTE Crab total spectrum connects smoothly to the 0.3-10 keV Crab total spectrum derived by \citet{willingale01}.
The HEXTE Crab total spectrum can best be described by a power-law model with an energy dependent photon index as given below:
\begin{equation} F_{\gamma} = 1.5703(14) \times (E_{\gamma}/0.06355)^{-2.097(2)-0.0082(16)\cdot \ln({E_{\gamma}/0.06355)}}
\label{eq_2} \end{equation}
 In Eq.\ref{eq_2}, $F_{\gamma}$ is expressed in ph/(cm$^2$\,s\,MeV) and $E_{\gamma}$ in MeV. 
Up to $\sim 180-200$ keV the extrapolation of the spectrum given by \citet{willingale01} overlaps the HEXTE spectrum, and thus provides
a proper representation for the ISGRI energy range we are effectively dealing with.

Therefore, our method for deriving ISGRI spectra enables us to determine the 20-300 keV AXP spectra without detailed knowledge of the 
instrument energy response, which at the time of our analysis still contained significant uncertainties. 
Thus, we derive from these spatial analyses total source spectra (sum of pulsed and unpulsed components). 

For the construction of spectra for the pulsed components we extracted count rates from the phase distributions
similarly as was done for the RXTE light curves. For the conversion to pulsed fluxes, also in this case, the Crab pulsed 
signal in the HEXTE 15 - 250 keV range (obs. id. 40805) was used as a reference source. This pulsed HEXTE spectrum is properly described by:
\begin{equation} F_{\gamma} = 0.4693(21) \times (E_{\gamma}/0.04844)^{-1.955(7)-0.0710(78)\cdot \ln({E_{\gamma}/0.04844)}}
\label{eq_3} \end{equation}
over the entire 15-250 keV energy range. This model has been verified on the pulsed ISGRI spectrum of PSR B1509-58 data, yielding
consistent results as reported in \citet{kuiper99}.

\subsection{COMPTEL spatial analysis} 
During the long mission lifetime of CGRO (April, 1991 -- May, 2000), most of the sky, 
particularly the Galactic Plane, has been viewed with long exposures. We used the exposure
accumulated for each source over the total mission duration, amounting for \axprxs\ 5.3 Ms, 
\axpu\ 4.2 Ms, \axpe\ 4.9 Ms, \axpee\ 4.8 Ms, and for \axp\ 4.2 Ms.  
Skymaps and source parameters can be derived with the maximum 
likelihood method, which is implemented in the standard COMPTEL data analysis package. 
See for the implementation of the maximum likelihood method for data from a 
Compton telescope \citep{deboer92}, and for the specific treatment of the instrumental
background structure in the COMPTEL data space \citep{bloemen94}. For the analysis 
and data selections we followed the approach described by \citet{zhang04}.  
Standard energy intervals were selected for the analysis: 0.75 -- 3 MeV,  3 -- 10 MeV and 
10 -- 30 MeV. For each of these energy intervals and for each AXP the source flux 
(or upper limit) was measured at the known source position in the maximum likelihood skymaps.


\section{\axprxs}

\axprxs\ was discovered in 1996 during the ROSAT (0.1-2.4 keV) all sky survey. 
X-ray pulsations at a 11-s period were subsequently detected with ASCA \citep{sugizaki97}. 
Its 0.8-10 keV spectrum turned out to be very soft and its general X-ray 
properties pointed to an AXP membership \citep[see e.g.][]{israel99a,kaspi99}. From regular RXTE PCA 
monitoring observations performed between Jan. 1998 and June 1999 a phase-coherent timing solution was obtained 
by \citet{kaspi99}. This demonstrated a high level of rotation stability. Since these early monitoring observations, 
however, the source experienced two glitches - one in Sept./Oct. 1999 and a second in April 2001 - each with 
different recovery behaviour \citep{kaspi00,kaspi03a,dallosso03}. For the period between the two glitches 
\citet{gavriil02} presented a phase-coherent timing solution with a positive $\ddot\nu$ indicative for a 
long-term glitch recovery. 

The morphology of the X-ray pulse profile of \axprxs\ is changing as a function of energy \citep[e.g.][]{sugizaki97,israel01,gavriil02}. Phase-resolved spectral analyses indeed showed significant spectral variations with pulse-phase, most 
pronounced in the photon power-law index \citep[e.g.][]{israel01,rea03,rea05}. 
Furthermore, the total phase-averaged unabsorbed 0.5-10 keV X-ray flux and photon spectral index appear to be time 
variable in a correlated way, with maximal fluxes and hardest spectra near the two glitch epochs \citep{rea05}. 

At optical/IR wavelengths 2 potential counterparts were identified within the Chandra $0\farcs7$ HRC-I error circle \citep[see e.g.]
[for more details]{israel03,harb05}. A search for radio emission at 1.4 GHz from \axprxs\ only yielded a $5\sigma$ upper limit of 
3 mJy at the position of the AXP \citep{gaensler01}.

Given the softness of the 0.5-10 keV X-ray spectra, the INTEGRAL detection reported by \citet{revnivtsev04} of a point 
source at the position of \axprxs\ between 18 and 60 keV was a big surprise. Below we will present in detail the new
high-energy characteristics of this AXP derived in this work: a) the discovery of the pulsed emission above $\sim 10$ keV (profiles, spectra)
using RXTE PCA/HEXTE and IBIS ISGRI data; b) ISGRI and COMPTEL results on the total emission.

\begin{figure}[t]
\centerline{\includegraphics[width=8.0cm,height=12.cm,bb=150 195 440 610]{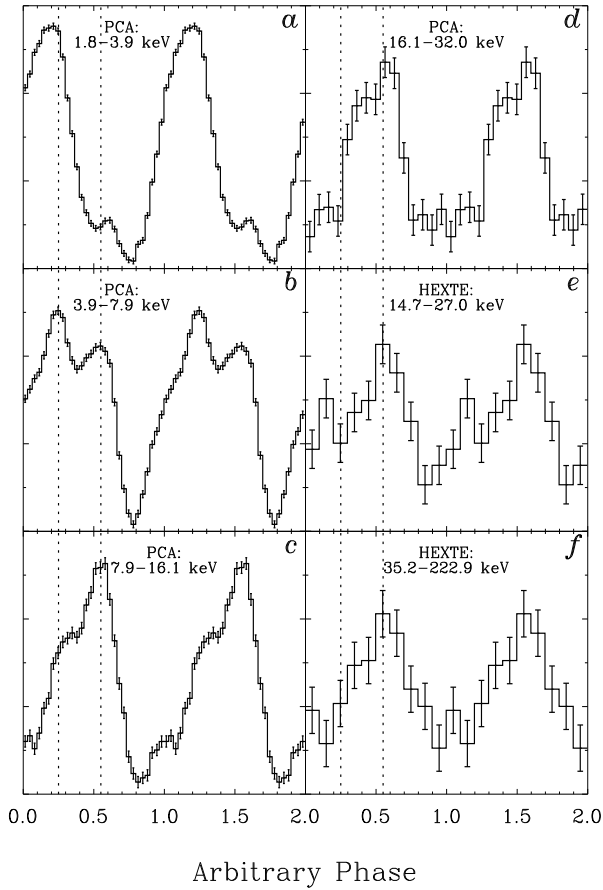}}
\caption[]{\label{pca_rxs_stack}RXTE PCA/HEXTE pulse profiles of \axprxs\ 
             for energies in the range 1.8-222.9 keV combining data collected between
             12 Jan 1998 and 26 Oct 2003 (see Table \ref{obs_table}). 
             Two cycles are shown for clarity. The vertical dotted lines at phases 0.25 and 
             0.55 serve as a guide to the eye for alignment comparisons. Note the drastic morphology changes with energy.}
\end{figure}

\subsection{\axprxs\ timing characteristics}

\subsubsection{RXTE PCA/HEXTE pulse profiles}
\begin{figure}[t]
\includegraphics[width=8.0cm,height=8.cm,bb=100 200 500 630]{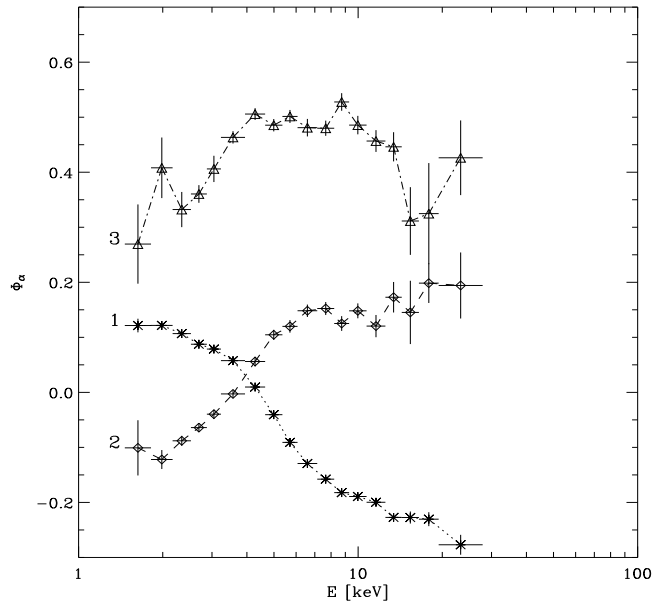}
\caption[]{\label{pca_rxs_fourier} Phase angles as a function of energy for the first 3 
             harmonics used in the truncated Fourier series fit (see Eq. \ref{eq_1}) of the RXTE PCA pulse profiles of \axprxs.
             The harmonics are labeled with their corresponding number.}
\end{figure}

Applying the timing analysis procedures outlined in Sect. \ref{rxte_timing} to the full set of RXTE observations of \axprxs\ listed
in Table \ref{obs_table}, resulted in a compilation of high-statistics {\em time-averaged} PCA/HEXTE pulse profiles for energies 
between $\sim 2-220$ keV (see Fig. \ref{pca_rxs_stack}). The ephemerides used in the folding/correlation process (see Sect. 
\ref{rxte_timing}) are given in \citet{kaspi00,gavriil02,kaspi03a}. 
For the first time pulsed emission is detected above $\sim 10$ keV: 
the non-uniformity significance of the 16.1-32.0 keV PCA pulse phase distribution (see Fig.\ref{pca_rxs_stack}d) is $14.2\sigma$ 
applying a $Z_2^2$-test and the HEXTE 35.2-222.9 keV profile (see Fig.\ref{pca_rxs_stack}f) deviates from uniformity at a $5.2\sigma$ level. 
Above 35.2 keV the significances in the HEXTE 35.2-64.1 and 74.3- 222.9 keV bands (the intermediate energy window with a large instrumental background feature has been omitted) are both $3.75\sigma$. 
Drastic morphology changes with energy are visible. The decomposition of the pulse profiles in terms of a finite number of 
harmonics (see Eq. \ref{eq_1}) provides a means to visualize a change in morphology with energy. 
The power ($a_k^2+b_k^2$, see Eq.\ref{eq_1}), derived from the time-averaged PCA pulse profiles of \axprxs, in the first harmonic 
is dominant over the power in the second and third harmonics, and the power in harmonics with $k \ge 4$ can be neglected 
\citep[see also][]{gavriil02}.
From Eq. \ref{eq_1} one can define a phase angle $\Phi_{\alpha}^k = \arctan(a_k/b_k)$ for each harmonic $k$. The energy dependence of $\Phi_{\alpha}^k$ for the first three harmonics is shown in Fig. \ref{pca_rxs_fourier}. 
For each harmonic it reveals a very smooth variation of the phase angle with energy. The shape of the profile is changing drastically 
between 2 and 10 keV. For energies above $\sim 15$ keV the phase angles for the 3 considered harmonics seem to converge to constant 
values, a necessary condition for stable pulse shapes.

\subsubsection{INTEGRAL IBIS ISGRI pulse profiles}

\begin{figure}[t]
\centerline{\includegraphics[height=8cm,width=6cm,angle=0,bb=170 195 420 605]{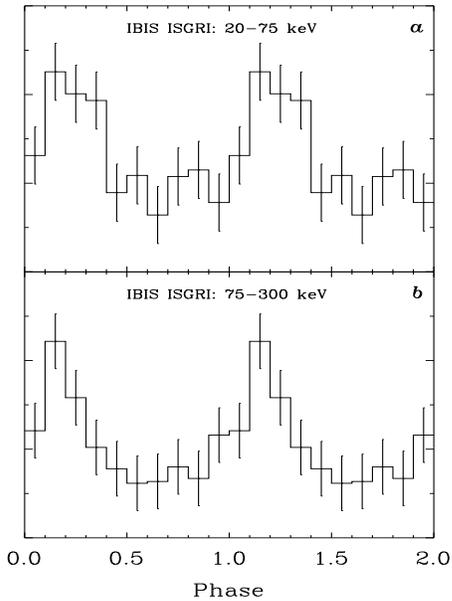}}
    \caption{\label{rxs_int_prof} INTEGRAL IBIS ISGRI pulse profiles of \axprxs\ for two energy ranges.
             The non-uniformity significances are $4.3\sigma$ and $3.6\sigma$ for 20-75 keV and 75-300 keV,
             respectively. Pulse maxima are found near phase $\sim 0.2$, corresponding to phase 0.55 in 
             Fig. \ref{pca_rxs_stack}.
            }
\end{figure}

We also performed a timing analysis for \axprxs\ using IBIS ISGRI data. Data from science windows taken during INTEGRAL
Revs. 36-120 satisfying our $14\fdg5$ off-axis constraint were included (effective on-axis exposure after screening $\sim$ 1,360 ks). 
The processing followed the guidelines presented in Sect. \ref{sect_int_timing} using the \axprxs\ ephemeris generated from
RXTE monitoring observations, given in Table \ref{eph_table}.
In the integral 20-300 keV ISGRI band we obtained a non-uniformity significance of $5.9\sigma$ applying a Z$_2^2$ test, which is 
comparable to the HEXTE result. In differential energy bands we found: 20-75 keV, $4.3\sigma$, and 75-300 keV $3.6\sigma$ 
(see Fig. \ref{rxs_int_prof} for the corresponding pulse profiles). The HEXTE and the ISGRI profiles above 75 keV are very similar, 
and suggest that the hard-X-ray \axprxs\ profile exhibits less structure than found below 10 keV. From these initial ISGRI timing 
results it is clear that highly significant profiles can be expected in the near future when significantly more IBIS ISGRI data on 
this source become available. 

\subsection{\axprxs\ spectral characteristics}

In this section we present new high-energy spectral information above 2.5 keV up to 30 MeV for \axprxs: 
(a) (time-averaged) pulsed emission from RXTE PCA and HEXTE; (b) pulsed emission from INTEGRAL IBIS ISGRI; 
(c) total (pulsed and unpulsed) emission from ISGRI and upper limits to the total 
emission from CGRO COMPTEL. Finally, the new spectra are compared with spectra reported
earlier for energies below 10 keV \citep[][for BeppoSAX LECS/MECS 0.4-10.8 keV and XMM Newton MOS/PN 0.5-10 keV, respectively]{rea03,rea05}.

\subsubsection{RXTE PCA/HEXTE pulsed spectrum}

The spectral procedures employed for the RXTE PCA and HEXTE data (see Sect. \ref{rxte_spectral}) resulted in a high-statistics
determination of the spectrum of the time-averaged pulsed emission of \axprxs\ in the $\sim 2.5-220$ keV energy range.
The PCA (aqua) flux values are derived assuming an absorbed double power-law spectral model, and are shown in 
a $\nu F_{\nu}$ representation in Fig. \ref{HE_SPECTRUM_RXS}. Also drawn is the best fitting spectral model to the PCA data points 
(2.5-36.9 keV; $\chi_{r}^2=1.11$ for 12 degrees of freedom; dashed line). The assumed absorbing Hydrogen column 
density $N_H$ in the spectral fit was $1.36\times 10^{22}$cm$^{-2}$ \citep{rea03}.   
The two power-law components become equally strong at $E_{cross}= 21.7 \pm 2.4$ keV: below this energy the power-law component
with index $\Gamma_1=2.60\pm 0.01$ dominates and above a component with very hard spectrum, index $\Gamma_2=-0.12\pm 0.07$.
It is clear that the pulsed spectrum hardens dramatically above 20 keV, however, the spectrum has to soften considerably to be consistent
with the index of $1.01(12)$ derived from a combination of PCA, HEXTE and IBIS ISGRI pulsed flux measurements for energies $\ga 15$ keV 
(see Sect.\ref{sect_ii_rxs_tm}). The HEXTE flux values (simultaneously derived) are fully consistent with the upturn found in the PCA 
spectrum (see Fig.\ref{HE_SPECTRUM_RXS}, blue datapoints).
 
\subsubsection{INTEGRAL IBIS ISGRI pulsed spectrum}
\label{sect_ii_rxs_tm} 
\begin{figure}[t]
\centerline{\includegraphics[height=8cm,width=8cm,angle=0]{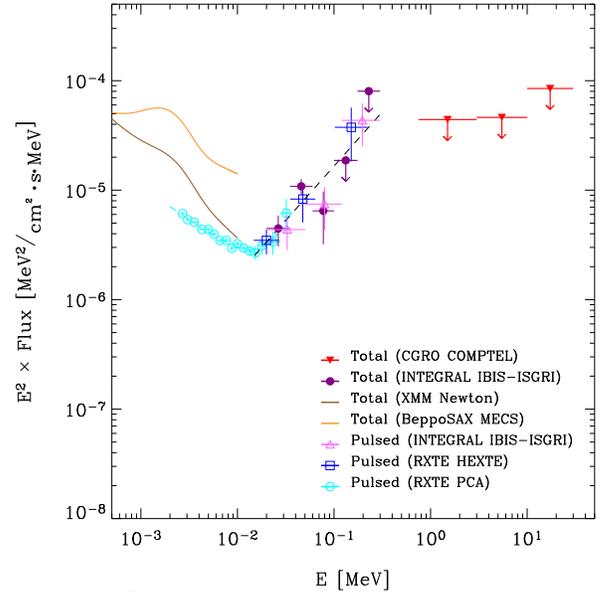}}
\caption{\label{HE_SPECTRUM_RXS}A $\nu F_{\nu}$ spectral representation of the total and pulsed high-energy 
             emission from \axprxs. The aqua (PCA), blue (HEXTE) and magenta (IBIS ISGRI) data points show the 
             {\em time-averaged} 2.5-300 keV pulsed spectrum. The black dashed line shows the best power-law model 
             fit to the combination of PCA, HEXTE and ISGRI pulsed flux values for energies above $\sim 15$ keV.
             The other measurements refer to the total emission spectrum: 0.5-10 keV, BeppoSAX LECS/MECS and XMM Newton 
             spectral models at different epochs (Rea et al. 2003,2005); 20-300 keV time-averaged (Revs. 36-106)
             IBIS ISGRI spectrum; and 0.75-30 MeV, time-averaged CGRO COMPTEL $2\sigma$ upper-limits. 
             Note the drastic hardening of the pulsed spectrum near 20 keV. The COMPTEL upper limits require another 
             spectral break somewhere between 300 and 750 keV.}
\end{figure}

Pulse profiles for \axprxs\ have been generated in three energy intervals
from 20 to 300 keV using IBIS ISGRI data from Revs 36-120. The pulsed (excess) count rates were determined 
and transformed to pulsed flux values, calibrated on the pulsed Crab spectrum. The three flux values
are also included in Fig. \ref{HE_SPECTRUM_RXS}. They are consistent with the data points derived 
from RXTE HEXTE for the same energy window, but measured at different epochs. 
Within the present statistical accuracies there is no indication for long-term flux variability above 20 keV, contrary 
to the reported strong flux variability for energies below 10 keV, clearly visible in Fig. \ref{HE_SPECTRUM_RXS}.
A power-law fit through the PCA, HEXTE and IBIS ISGRI pulsed flux measurements for energies above 15 keV yielded a
photon index of $1.01\pm 0.12$. This model is shown in Fig. \ref{HE_SPECTRUM_RXS} as a dashed black line for the 15-300 keV energy
range. There is no indication yet for a second spectral break up to 300 keV. 
  
\subsubsection{INTEGRAL IBIS ISGRI and COMPTEL total spectrum}

\axprxs\ was detected at a $6.5\sigma$ level in the 18-60 keV energy band by \citet{revnivtsev04} using IBIS ISGRI data
from a 2 Ms ultra deep INTEGRAL survey of the Galactic center region performed in Aug. - Sept. 2003 (INTEGRAL Revs. 104-107 \& 111-113).
Its 18-60 keV flux was $2.2\pm 0.3$ mCrab. In this work we analyzed (almost independent) IBIS ISGRI data from all publicly available 
INTEGRAL observations performed between revolutions 36 and 106 in which the angular distance between \axprxs\ and the science window 
pointing was $\le 14\fdg5$. The effective on-axis exposure after screening is about 974 ks. 
Mosaic images have been made using OSA 4.1 procedures for 5 broad differential energy bands, 20-35, 35-60, 60-100, 100-175 and 175-300 keV,  
in order to determine the total hard X-ray/soft $\gamma$-ray spectrum of this AXP. \axprxs\ was detected significantly only in the 20-35 
and 35-60 keV energy bands with flux levels consistent with the 18-60 keV flux measurement by 
\citet{revnivtsev04}. The resulting IBIS ISGRI spectral measurements are shown as purple data points in Fig. \ref{HE_SPECTRUM_RXS}.
Unfortunately, the statistics in these IBIS ISGRI data points are still too poor to constrain the underlying emission model in the 
20-300 keV range: the fit quality specified by a reduced $\chi^2_r$ of $1.77$ for 3 d.o.f. assuming a simple power-law 
model is still acceptable; the resulting photon index is $1.44 \pm 0.45$, slightly softer than, but consistent with, the pulsed flux 
spectrum derived from the combined fit to the PCA, HEXTE and IBIS ISGRI pulsed flux measurements at $\ga 15$ keV. 

Finally, at energies above 750 keV till 30 MeV we generated skymaps using all CGRO COMPTEL observations with \axprxs\ in its field of 
view, spread over its full 9 year mission lifetime.  The source was not seen for any of the standard energy intervals, and we derived 
($2\sigma$) flux upper limits, also shown in Fig. \ref{HE_SPECTRUM_RXS}. 

\subsubsection{Discussion of spectra and pulsed fraction of \axprxs}

Comparing the total IBIS ISGRI spectrum derived in the spatial analysis with the pulsed time-averaged spectra 
from RXTE PCA/HEXTE and ISGRI, we see that the \axprxs\ emission is consistent with about 100\% pulsation 
for energies in excess of 20 keV. For energies below 20 keV, the situation is very different. To highlight this, we included
in Fig.\ref{HE_SPECTRUM_RXS} the emission models (best spectral fits) for the total 0.5-10 keV X-ray spectra measured at two 
different epochs \citep[][for BeppoSAX LECS/MECS 0.4-10.8 keV and XMM Newton MOS/PN 0.5-10 keV, respectively]{rea03,rea05}.
Given the apparent drastic spectral change in both shape and normalization for energies below 10 keV we cannot uniquely quantify 
the pulsed fraction as a function of energy comparing with the time-averaged high-statistics PCA pulsed flux measurements.  
Obviously, the BeppoSAX measurement during August 2001 does not connect to the RXTE/INTEGRAL spectra. \axprxs\ was apparently 
in a (very) different high state during the BeppoSAX observation, which took place in the recovering phase from the secondly 
reported glitch \citep[see][]{rea05}. 
The XMM Newton spectrum connects very smoothly to the total spectrum measured by INTEGRAL, and suggests a variation
in pulsed fraction from $\sim$ 25\% at 2 keV to 100\% at 20 keV. Our time averaged results for the total and pulsed emission  
above 20 keV suggest a very stable behaviour of this AXP at these higher energies. It seems that a stable and very hard pulsed 
component is dominating the emission at these energies.
Interestingly, the COMPTEL MeV upper limits then require a spectral break in this hard spectrum of \axprxs\ somewhere 
between 300 and 750 keV.


\section{\axpu}

\axpu\ was discovered in the early seventies by the scanning {\it Uhuru\/} X-ray observatory 
\citep[see e.g.][]{forman78}. The X-ray sky survey performed by the SSI (1.5-20 keV) on {\it Ariel V\/} 
\citep{warwick81} and X-ray observations by SAS-3 and HEAO-1 confirmed the source and refined
its position considerably \citep[see e.g.][for an X-ray position with $23\arcsec$ accuracy]{reid80}. 
Data from the non-imaging ME collimator ($45\arcmin$ FWHM) aboard EXOSAT revealed a coherent 25 minute 
signal, which was later absent in two shorter EXOSAT observations \citep{white87}. These authors also presented 
more accurate ($3\farcs2$) positional information for \axpu\ using both EXOSAT LEIT and {\it Einstein\/} (HEAO-2) 
HRI data and demonstrated the lack of a bright optical counterpart, ruling out the presence of a massive companion. 

The discovery in ROSAT all-sky survey data of a Be/X-ray binary, RX J0146.9+6121, by \citet{motch91} at only $24\arcmin$ from 
\axpu\ showed that there had been source confusion. Revisiting the EXOSAT 1985 data, \citet{israel94} discovered a 8.7 s 
periodicity, only detectable in the 1-3 keV energy band. 
Finally, using ROSAT PSPC data, \citet{hellier94} unambiguously demonstrated that the 8.7 s periodicity originated from \axpu\ and the 25 min 
signal from the Be/X-ray binary RX J0146.9+6121, which moreover showed transient activity.

A good quality phase-averaged 0.5-10 keV X-ray spectrum of \axpu\ was derived by \citet{white96} analyzing ASCA SIS/GIS data.
The physically most pausible model consists of an absorbed black-body plus power-law component with best-fit model parameters: black-body 
temperature $0.386(5)$ keV; photon-index $3.67(9)$ and N$_{\hbox{\scriptsize H}}$ $9.5(4)\times 10^{21}$ cm$^{-2}$. These authors, using data covering a much wider time baseline, also confirmed the spin-down timescale of $\sim 1.2\times 10^5$ year, reported earlier by 
\citet{hellier94}. The spin-down energy release is much lower than the observed X-ray luminosity, thereby excluding \axpu\ to be a spin-down 
powered pulsar. 

RXTE monitoring observations since November 1996 \citep{gavriil02}  proved that \axpu\ is a very stable rotator, like spin-down powered pulsars, and
ASCA data taken in a 2 year gap of RXTE observations seemed to indicate a pulsar-like glitch in its rotation behaviour
\citep{morii05}. Time-averaged RXTE PCA pulse profiles in two different energy bands showed significant morphology changes with energy, 
consistent with earlier results of much lower significance \citep{israel94,white96,israel99b}.

Recent X-ray observations of \axpu\ by Chandra \citep{juett02,patel03} and XMM \citep{gohler05} improved not only the positional 
accuracy down to $0\farcs5$, but also provided high-quality (phase-resolved) spectra which are mutually consistent and
appeared featureless. The phase-resolved spectra showed significant changes in photon-index and black body temperature as a function 
of pulse phase.

At lower energies, \citet{hulleman00} discovered a faint optical counterpart within the \axpu\ {\it Einstein\/} X-ray error circle, followed by
the detection of optical pulsations with a high pulsed fraction from this counterpart \citep{kern02}. No radio counterpart has been found yet 
\citep{gaensler01}.

At higher energies the discovery was reported by \citet{hartog04} of hard X-rays/soft $\gamma$-rays (20 -- 150 keV) in INTEGRAL IBIS ISGRI data  
of a deep 1.6 Ms observation of the Cassiopeia region. A detailed analysis of this long observation \citep{hartog06}, proved that the spectrum in this energy range is very hard with power-law photon index $\Gamma=0.73 \pm 0.17$. The X-ray luminosity between 20 and 100 keV was found to be 
$5.9 \times 10^{34}$ erg s$^{-1}$, a factor 440 higher than the rotational energy loss. They also reported flux upper limits from COMPTEL 
between 0.75 and 30 MeV, which indicated that the hard spectrum has to break/bend between $\sim$ 100 keV and 0.75 MeV. 

In this work we will present for \axpu:  a) for the first time the PCA/ HEXTE timing and spectral results for energies above $\sim 8$ keV 
using all publicly available RXTE data, b) timing and spectral results from analysis of ASCA GIS 0.5 - 10 keV data, and c) a spectrum with 
higher statistical accuracy of the total emission as seen by IBIS ISGRI over the 20-300 keV energy range using all available public, 
Core Program (Galactic plane scans; 4U 0115+63 Target of Oppotunity Observation) and Open Time (Cassiopeia region; IGR J00291+5934 TOO) 
INTEGRAL data.

\subsection{\axpu\ timing characteristics}

\begin{figure}[t]
\centerline{\includegraphics[height=8.5cm,width=10cm,angle=90,bb=55 172 515 670]{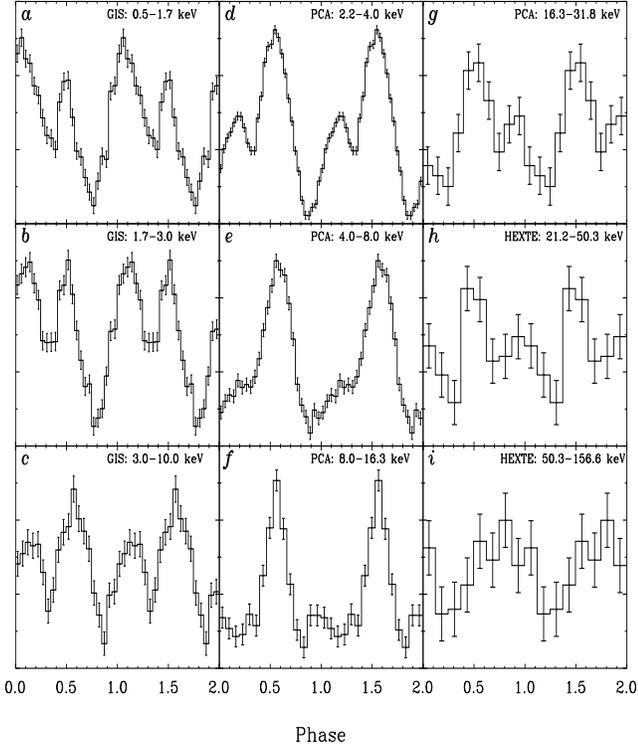}}
    \caption{\label{pca_u_stack} ASCA GIS, RXTE PCA and HEXTE pulse profiles of \axpu\ 
    for energies in the range 0.5-156.8 keV. Two cycles are shown for clarity. The RXTE profiles shown in panels d--i are
    time-averaged and based on observations performed between March 1996 and September 2003. The GIS profiles (panels a--c) are
    from ASCA observations performed in July/August 1999 totaling $\sim 120$ ks of exposure. For the first time significant 
    deviations from uniformity are visible at energies $> 8$ keV: $Z_2^2$ is $16.6\sigma$ and $5.9\sigma$ for the PCA 
    8.0-16.3 and 16.3-31.8 keV bands, respectively; the significances for the HEXTE profiles are $3.4\sigma$ and $2.0\sigma$ 
    for the 21.2-50.3 and 50.3-156.6 keV bands. Note the drastic morphology changes with energy.
           }
\end{figure}

Time-averaged RXTE pulse profiles of \axpu\ are shown in Fig. \ref{pca_u_stack} for energies between 2.2 and 156.6 keV (PCA/HEXTE). 
We used in the folding/correlation process (see Sect. \ref{rxte_timing}) the ephemerides given by \citet{gavriil02}. 
The 2.2-4.0 keV PCA profile (panel d) shows two distinct pulses near phases 0.2 and 0.55. Moving up in energy the pulse near 0.2 
loses significance and is gone for energies above $\sim 8$ keV (panels e--f). Instead above $\sim 8$ keV a feature near phase 
0.9 pops up, which is visible up to $\sim 50$ keV (panels f--h). 
Above 50 keV (panel i) the phase region between the two pulses at 0.55 and 0.9 seems to be filled in by a new component, but the 
statistics is too poor to make stringent conclusions. 

Most importantly, we detect for the first time significant pulsed emission from \axpu\ at energies above $\sim 8$ keV: $16.6\sigma$ 
and $5.9\sigma$ for the PCA 8.0-16.3 and 16.3-31.8 keV bands, respectively.
To investigate further the increase in strength of the pulse near phase 0.2 towards lower energies, we extended the energy baseline 
by including also our results from a timing analysis of ASCA GIS (0.5-10 keV) data from observations performed in July/August 1999 
($\sim 120$ ks exposure). Three profiles (energy bands 0.5-1.7, 1.7-3.0 and 3.0-10 keV) are shown in panels a--c of Fig.\ref{pca_u_stack}. 
Indeed, for the pulse near phase 0.2 the trend of increasing strength towards lower energies is continued in the 
ASCA GIS 0.5-1.7 and 1.7-3.0 keV energy ranges. 

\begin{figure}[t]
\includegraphics[width=8.0cm,height=8.cm]{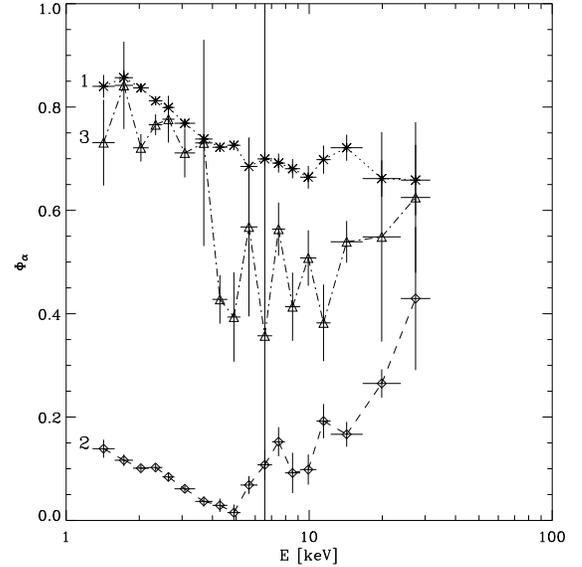}
\caption{\label{pca_u_fourier} Phase angles as a function of energy for the first 3 
   harmonics of the truncated Fourier series fit of the RXTE PCA pulse profiles of \axpu.
   Note the sign reversal of the energy derivative of $\Phi_{\alpha}$ for the second harmonic near 5 keV.}
\end{figure}

A decomposition of the PCA pulse profiles in terms of 5 Fourier harmonics shows that above 2 keV most of the signal power is  
embedded in the first harmonic which becomes more and more dominant over the power in the second harmonic, the next powerful harmonic, 
with increasing energy.
Below $\sim 2$ keV the power in the second harmonic is dominant over that in the first. The behaviour of
the phase angles $\Phi_{\alpha}^k$ as a function of energy for the first three harmonics is shown in Fig. \ref{pca_u_fourier}.
Similar to \axprxs\, but with different trends, also now a smooth energy dependency of $\Phi_{\alpha}^k$ is 
shown for each harmonic with significant power. 
The exception is the derivative of $\Phi_{\alpha}^2$ to energy which flips sign near 5 keV.  For \axpu\ it is less obvious 
that the pulse profile remains stable above 20 keV, than seems to be the case for \axprxs.

\subsection{\axpu\ spectral characteristics}

In this section we present: a) the {\em time-averaged} pulsed spectrum (2.2-102 keV) of \axpu\ based on RXTE PCA and 
HEXTE measurements. The pulsed spectrum is extended down to 0.8 keV by including our spectral results from the ASCA 
GIS observations ($\sim$120 ks); b) the {\em time-averaged} total (pulsed and unpulsed) spectrum 
from our analysis of INTEGRAL IBIS ISGRI skymaps (20-300 keV), in comparison with that from Chandra ACIS-S CC-mode 
data \citep[0.5-7 keV;][]{patel03} and the CGRO COMPTEL (0.75-30 MeV) upper limits \citep{hartog06}.

\subsubsection{RXTE PCA/HEXTE and ASCA GIS pulsed spectrum}

\begin{figure}[t]
\centerline{\includegraphics[height=8cm,width=8cm,angle=0]{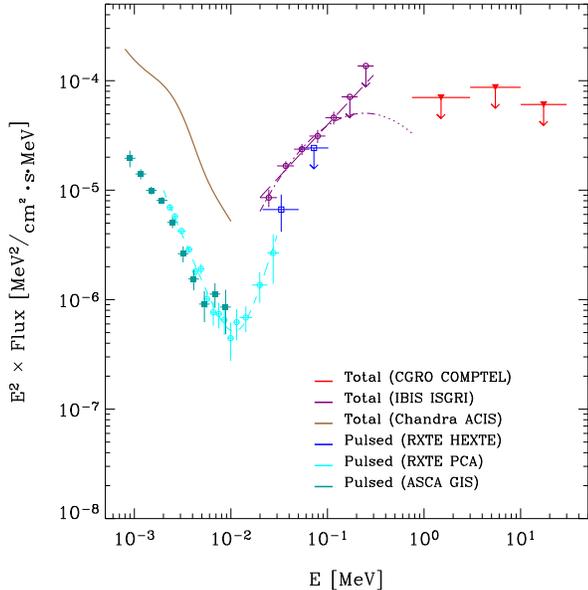}}
\caption{\label{HE_SPECTRUM_U} A $\nu F_{\nu}$ spectral representation of the total and pulsed high-energy emission
  from \axpu. The aqua, blue and dark cyan data points/curves represent the pulsed emission component of the spectrum 
  (0.8-102 keV) based on RXTE PCA/HEXTE observations (time-averaged) and ASCA GIS. Note the dramatic hardening of the pulsed 
  spectrum near 10 keV. The other measurements refer to the total emission spectrum: 0.8-10 keV, Chandra ACIS \citep{patel03}, 
  20-300 keV, INTEGRAL IBIS ISGRI (this work) and 0.75-30 MeV, CGRO COMPTEL \citep{hartog06}. The COMPTEL $2\sigma$ upper 
  limits require a spectral break somewhere between 140 and 750 keV.
  }
\end{figure}

Assuming an underlying absorbed \citep[$N_H=9.3\times 10^{21}$cm$^{-2}$;][]{patel03} double power-law photon model 
we fitted the pulsed PCA excess counts (2.2-31.5 keV) in a forward folding procedure (see Sect. \ref{rxte_spectral}),
resulting in a high-statistics time-averaged pulsed spectrum. 
The best model fit ($\chi_{r}^2=0.7$; d.o.f. 15 - 4) yielded for the soft and hard photon indices $\Gamma_1 = 4.09 \pm 0.04$ 
and  $\Gamma_2 = -0.8^{+0.10}_{-0.07}$, respectively. This is shown as a dashed aqua curve in Fig. \ref{HE_SPECTRUM_U}, together
with the ``deconvolved'' unabsorbed flux values (aqua colored open circles). The power-law model components become equally strong 
at $11.46\pm 0.64$ keV. 

The striking spectral hardening of the pulsed emission of \axpu\ around 10 keV is dramatic, much more pronounced than the 
hardening observed for \axprxs.

The HEXTE pulsed flux values (Fig. \ref{HE_SPECTRUM_U}; dark-blue open squares) are in line with the extrapolation of the PCA 
hard-power-law model component.

The ASCA GIS 0.8-10 keV pulsed flux values which we derived from the July/August 1999 observations are also shown in Fig. 
\ref{HE_SPECTRUM_U} (filled squares) and are within the systematic/statistical uncertainties consistent with the 
time-averaged PCA fluxes in the overlapping energy range. This suggests a rather stable pulsed spectrum both in 
shape and normalization.   

\subsubsection{INTEGRAL IBIS ISGRI total spectrum}

\citet{hartog06} report the detection and total spectrum of \axpu\ in the 20-100 keV energy range 
analyzing IBIS ISGRI data from INTEGRAL observations of the Cassiopeia region performed in Dec. 2003 
(Revs. 142 - 148). In this work all available INTEGRAL data (open time/core program) 
have been used from observations made between 3 March 2003 (Rev. 47) and 28 December 2004 (Rev. 269) in 
which the source was within $14\fdg5$ from the pointing axis.
The total screened exposure time for the set of accumulated science windows is about 2 Ms, which corresponds 
to an effective on-axis exposure of about 858 ks. Image mosaics have been generated using OSA 4.1 software in
7 differential energy bands, 20-30, 30-45, 45-65, 65-95, 95-140, 140-205 and 205-300 keV. In these (time-averaged) maps\footnote{Be X-ray binary RX J0146.9+6121 at only $24\arcmin$ from \axpu\ was detected, fully resolved, 
only in the first three energy bands.}
\axpu\ is clearly detected up to and including the 95-140 keV energy range ($7.8\pm 1.0$ mCrab). The 20-300 
keV flux measurements are shown as purple data points in Fig. \ref{HE_SPECTRUM_U}. A simple power-law fit to 
these flux points yielded a hard photon index of $1.05\pm 0.11$ ($\chi^2_{r}$ = 0.86 for d.o.f. 6 - 2; the purple 
dashed line in Fig. \ref{HE_SPECTRUM_U}). The difference with the value ($0.73 \pm 0.17$) reported for the index by \citet{hartog06} is 1.6$\sigma$.

\subsubsection{Discussion of spectra and pulsed fractions of \axpu}

The INTEGRAL spectrum of the total emission above 20 keV can be compared with the same Chandra spectrum for energies below 10 keV
\citep[see Fig. \ref{HE_SPECTRUM_U} the dark orange solid line;][]{patel03}. As is the case for the pulsed spectra above and below $\sim$ 10 keV,
these two spectra are drastically different, and together reveal a sharp minimum in luminosity of \axpu\ around 10 keV. 

Including now in this high-energy spectral picture the time-averaged CGRO COMPTEL 
$2\sigma$ flux upper limits (red triangles in Fig. \ref{HE_SPECTRUM_U}) from \citet{hartog06} 
confirms that the power-law model statisfactorily describing the 20-300 keV 
total spectrum will not extend into the MeV range, but must break somewhere 
between 140 and 750 keV. To investigate this further,  we also fitted a spectral model 
with an energy-dependent photon index, $F_{\gamma}=K\cdot E_{\gamma}^{-(\Gamma + \alpha \ln(E_{\gamma}))}$,
to the 20-300 keV IBIS ISGRI measurements (see Fig. \ref{HE_SPECTRUM_U} dashed 
dotted purple line). The model extrapolation towards the MeV energy range is consistent with the COMPTEL upper limits, 
but the improvement of the fit is insufficient to claim a change in spectral shape in the 20-300 keV window.    

Comparing now the total and pulsed high-energy spectra of \axpu, 
the Chandra total flux values below 10 keV are about 10 times 
higher than the time-averaged RXTE PCA and ASCA GIS pulsed flux measurements.
This is consistent with pulsed fraction estimates of $\sim$ 10\% reported
by \citet{patel03} and
\citet{gohler05} using solely Chandra and XMM Newton data, respectively.
The total spectrum measured by INTEGRAL above 20 keV with slope $1.05\pm 0.11$ 
is significantly softer than the pulsed spectrum measured by RXTE above
$\sim$ 10 keV with slope  $-0.8^{+0.10}_{-0.07}$. As a result, the pulsed
fraction appears to vary with energy from $\sim$ 10\% at 20 keV to
$\sim$ 100\% between 80 and 100 keV.


\section{\axpe}

Near the center of SNR CTB 109 (G 109.1-1.0) a strong compact X-ray source was found 
by \citet{gregory80}, later dubbed \axpe, which appeared to be an X-ray pulsar with 
a period of $3.489(2)$ s \citep{fahlman81}. 
Analyzing more IPC data, this period turned out to coincide with the first 
harmonic of a fundamental 6.978 s period \citep{fahlman83}. 

No radio enhancement at the position of the X-ray pulsar could be identified 
down to a level of 0.5 mJy \citep[WSRT, 21 cm;][]{hughes84}, 
0.2 mJy \citep[VLA, 20 cm;][]{gregory83} and even $50\mu$Jy \citep[VLA, 
20cm;][]{coe94,kaspi03}. 
On the other hand, the lack of a bright visual counterpart ($V\ge 21$) ruled out 
a supergiant or massive main-sequence star association 
\citep{fahlman81}.
 
Further X-ray observations of \axpe\ in the eighties with Tenma (Astro-B), 
EXOSAT and Ginga (Astro-C) \citep[see e.g.][]{koyama87,morini88,hanson88,koyama89,iwasawa92} revealed a 
steady spin-down, too slow to power the observed X-ray luminosity, and a very soft spectrum.
X-ray flux measurements from two different Ginga observations spread $\sim 8$ months apart
indicated considerable flux variations, accompanied with clear pulse morphology 
changes and probably a decrease in spin-down rate \citep{iwasawa92}. 
The X-ray picture was refined considerably in the nineties using data from 
BBXRT, ASCA (Astro-D), ROSAT, BeppoSAX and RXTE.
\citet{corbet95} showed for the first time that a two-component spectral model - 
black-body plus power-law - could fit 
the ASCA and BBXRT spectral data satisfactorily without invoking spectral (line) 
features. This finding was confirmed by \citet{rho97} analyzing
ROSAT PSPC, ASCA and BBXRT data simultaneously, and by \citet{parmar98} analyzing 
BeppoSAX LECS and MECS data.
RXTE timing measurements \citep{mereghetti98} further tied down earlier limits on 
the projected semi-axis to 0.03 lts, leaving room only for a white 
dwarf companion, helium-burning star with mass smaller than 0.8 M$_\sun$ or 
a main-sequence star viewed under very small inclination angles.

\citet{kaspi99} obtained for the first time a phase coherent timing solution 
for \axpe, indicating great stability (rms 0.01 cycles) over the 
2.6 year timespan (29-Sept-1996 -- 12-May-1999) of their RXTE monitoring 
observations. This stability makes binary accretion scenarios very 
unlikely and favours a magnetar interpretation. Additional RXTE monitoring data 
showed that this rotational stability was maintained throughout 
an extended 4.5 yr period, although the inclusion of 
$\ddot\nu$ was required in the timing model \citep{gavriil02}.
These authors also found that the pulse morphology did not change significantly 
with time \citep[cf.][who claimed changes in morphology with time]{iwasawa92}. 
Moreover, there was no evidence for large
variability in the pulsed flux in line with earlier work by \citet{baykal00} using 
RXTE data, but in contrast with the Ginga findings \citep{iwasawa92}. 
Furthermore, \citet{gavriil02} showed, convincingly for the first time, that 
the pulse profile morphology of \axpe\ changes
with energy \citep[cf.][for earlier indications]{hanson88}.

Observations with the Chandra X-ray observatory in January 2000 provided an 
X-ray position with subarcsecond accuracy \citep{hulleman01,patel01}. 
In the $0\farcs6$ error circle (99\%) a very faint ($K_s=21.7\pm 0.2$ mag) 
near-infrared counterpart was identified \citep{hulleman01}, 
excluding models in which the source is powered by disk accretion.  
The total X-ray (0.5-7.0 keV) spectrum of \axpe\ appeared featureless and could 
best be described by a combination of a black body $kT=0.412(6)$ keV 
plus power-law $\Gamma=3.6(1)$ absorbed by a Hydrogen column $N_H=(9.3\pm0.3) 
\times 10^{21}$ cm$^{-2}$ \citep{patel01}.

In June 2002 RXTE observed an outburst of \axpe\ during which both the pulsed and persistent X-ray 
emission increased by more than an order of magnitude relative to their quiescent levels \citep{kaspi03}. 
During the course of the observation a spectral softening occured, and a 
significant pulse profile change was observed. In the meantime
the pulsar underwent a sudden spin-up (glitch) followed by a large increase in 
spin-down rate lasting for more than 18 days.
Furthermore, 80 X-ray bursts were detected during the 14.4 ks RXTE observation 
with durations ranging from 2 ms to 3 s.
The outburst properties of \axpe\ share strong similarities with SGR outburst 
characteristics unifying thereby conclusively AXPs and SGRs
and supporting strongly a magnetar interpretation. More detailed information on 
the burst characteristics can be found in \citet{gavriil04}. 
\citet{woods04} presented a comparison of the X-ray emission characteristics of 
\axpe\ before, during and after the June 2002 outburst using
data from XMM-Newton and RXTE. They quantified the changes of the temporal and 
spectral properties and derived recovery timescales.
The X-ray flux increase and subsequent decay can be described by two distinct 
components: one component linked to the burst activity with a 
timescale of $\sim 2$ days and a second component which decays over the course 
of the year according to a power-law in time ($F \propto t^\alpha$) 
with index $\alpha = -0.22\pm0.01$. The latter component behaves similarly in 
time as the observed near-infrared flux decay \citep{tam04} and
thus these seem to be linked to a common physical mechanism likely acting in the 
pulsar's magnetosphere.

In this work we derived for the first time the time-averaged timing/spectral 
properties of \axpe\ for energies above $\sim 8$ keV using archival
RXTE PCA/HEXTE data and we compared the pulsed spectrum with the upper limits
from IBIS ISGRI and COMPTEL \citep{hartog06}.

\subsection{\axpe\ timing characteristics}

\begin{figure}[t]
\centerline{\includegraphics[height=12cm,width=8cm,angle=0,bb=125 155 465 650]{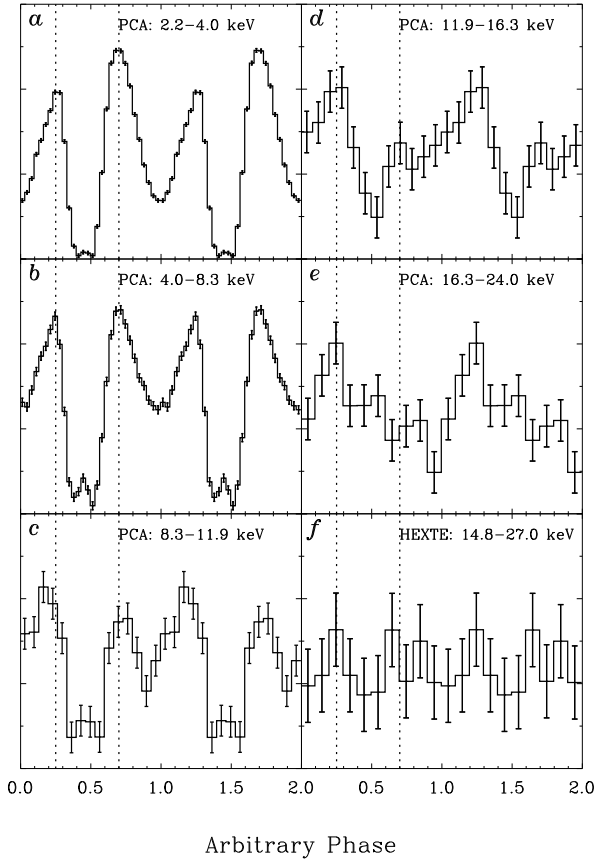}}
  \caption{\label{pca_e_stack}RXTE PCA/HEXTE pulse profile collage of \axpe\ 
  for energies in the range 2.2-27.0 keV combining data collected between
  29 Sept 1996 and 28 Oct 2003 (see Table \ref{obs_table}). 
  Two cycles are shown for clarity. The vertical dotted lines at phases 0.25 and 
  0.70 serve as a guide to the eye for alignment comparisons. Clear pulse profile 
  morphology changes with energy are present.
  Pulsed emission is detected up to $\sim 24$ keV.
  Note the enhancement within the main valley between the two peaks near phase 
  0.45, also visible in Fig. 7c of \citet{gavriil02}.}
\end{figure}

The {\em time-averaged} RXTE PCA/HEXTE pulse profiles combining all observations of \axpe\ listed in Table \ref{obs_table}, 
except the 14.4 ks period of outburst in 2002, are shown in Fig. \ref{pca_e_stack} for energies between 2.2 and 27 keV. 
In the folding/correlation process (see Sect. \ref{rxte_timing}) we used the ephemeris 
given by \citet{gavriil02}. The screened PCU-2 exposure amounts $\sim 747$ ks, while the corresponding 
dead-time corrected on-source HEXTE exposures are 256.4 and 267.7 ks for cluster-0 and 1, respectively. 
For the first time significant deviations from uniformity are visible at energies $\gtap 8$ keV: 
$Z_2^2$ is $10.6\sigma$, $5.2\sigma$ and $3.1\sigma$ for the PCA 8.3-11.9, 11.9-16.3 and 16.3-24 keV bands, respectively 
(Fig. \ref{pca_e_stack} panels c-e). The HEXTE sensitivity is still too low to detect the underlying pulsed emission 
above $\sim 15$ keV (Fig. \ref{pca_e_stack} panel f).

The morphology of the double-peaked pulse profile changes gradually moving up in 
energy from 2.2 to 24 keV: the dominant pulse near phase 0.7 
in Fig. \ref{pca_e_stack} at energies below $\sim 4$ keV loses significance with 
respect to the second pulse near phase 0.25. The latter
dominates at energies above $\sim 8$ keV. Also note the existence of a narrow 
pulse-like feature in the deepest minimum near phase 0.45, most
striking in panel b (4-8.3 keV) of Fig. \ref{pca_e_stack}. The latter feature
is also clearly visible in Fig. 7c of \citet{gavriil02}.

\begin{figure}[t]
\includegraphics[width=8cm,height=7.5cm,bb=90 200 515 645]{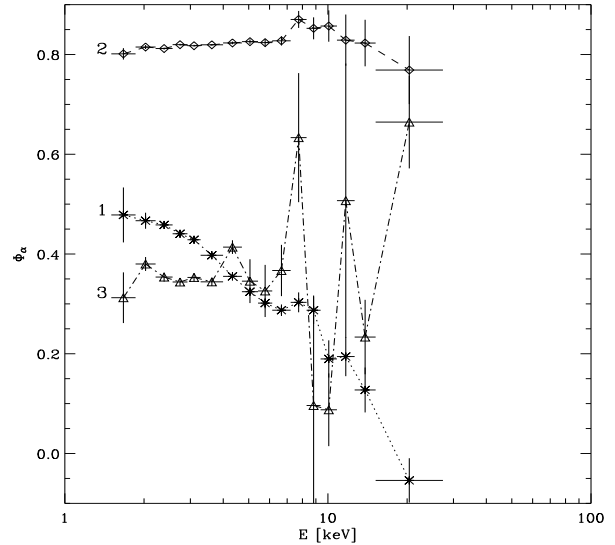}
\caption{\label{pca_pa_e2259} Phase angles as a function of energy for the first 3 
   harmonics used in the truncated Fourier series fit of the RXTE PCA pulse profiles of \axpe.
   Note that the position of the phase angle of the dominant second harmonic (labelled 2) is 
   more or less constant (energy independent), while that of the first harmonic (labelled 1) 
   decreases with increasing energy (shifts to the right).}
\end{figure}

More quantitative information on the energy dependency of the pulse morphology 
can be obtained through a decomposition of the pulse profiles
in Fourier components. At least 5 harmonics are required to adequately fit the 
profiles in the various energy slices, mainly driven by
the existence of the narrow, significant feature in the main valley between the 
peaks. This was also shown by \citet{gavriil02}. While the power of the second harmonic (2 maxima per cycle) 
dominates over that of the first up to $\sim 10$ keV (beyond which they become 
statistically equally strong) and the phase angle $\Phi_{\alpha}^2$ of 
the second harmonic remains more or less at the same position in phase, the 
first harmonic (broad component) shifts gradually to the right in 
phase with increasing energy (see Fig. \ref{pca_pa_e2259}). This energy dependent 
shift reflects the collapse at higher energies of the dominant pulse 
at low energies near phase 0.7.  

\subsection{\axpe\ spectral characteristics}

In this section the {\em time-averaged} pulsed spectrum of \axpe\ is 
derived based on the RXTE PCA \& HEXTE observations given in Table 
\ref{obs_table}. The high-energy picture has been completed by 
comparing our results with the total emission spectrum (2-10 keV) as determined by 
\citet{patel01} and with IBIS ISGRI (20-300 keV) and COMPTEL (0.75-30 MeV) 
upper limits \citep{hartog06}.

\subsubsection{RXTE PCA/HEXTE pulsed spectrum}
Employing the same counts-extraction technique as used above for the other AXPs, the pulsed 
excess counts from the PCA profiles are converted into flux values 
taking into account the different PCU exposures and sensitivities, and adopting 
an absorbing Hydrogen column of $N_H=(9.3\pm0.3) \times 10^{21}$ cm$^{-2}$ \citep{patel01}.

Assuming a simple absorbed power-law model the fit resulted in a poor $\chi^2$ of $26.24$ for
$14-2$ degrees of freedom ($\chi^2_r = 2.187$), indicating an inappropriate fitting function 
(there is a $\sim 1\%$ probability that such a high value of $\chi^2_r$ is obtained at random assuming
that the fit function is a good representation of the unkown parent function).
An absorbed double power-law model, however, yielded a reasonable $\chi^2$ of $16.70$ for $14-4$ degrees 
of freedom ($\sim 10\%$ random probability). The improvement $\Delta \chi^2$ of $9.55$ adding two additional fit parameters
translates to a $\sim 3\sigma$ improvement adopting a maximum likelihood ratio test. Thus, the double power-law model
provided a significant improvement in describing the same spectral data.
Adopting this double-power-law model as underlying photon model spectrum, the pulsed flux values are shown in Fig. 
\ref{HE_SPECTRUM_1E} as aqua colored filled circles, and the best fit double-power-law model
as a dashed aqua colored line.
 
The soft and hard power-law indices are $\Gamma_1=4.26\pm 0.01$ and $\Gamma_2=-1.02^{+0.24}_{-0.13}$, 
respectively, and the power-law-model components become equally strong at $E_{int}=15.8\pm 2.3$ keV. 
Thus, also for this AXP we witness the onset of a dramatic hardening of the pulsed spectrum beyond 
$\sim 10$ keV, although confirmation for energies beyond 20 keV is required.
For HEXTE the source is too weak to be detected; $2\sigma$ upper limits (dark blue) are shown in Fig. 
\ref{HE_SPECTRUM_1E}.    

\subsubsection{Discussion of spectra and pulsed fractions of \axpe}
\begin{figure}[t]
\centerline{\includegraphics[height=8cm,width=8cm,angle=0]{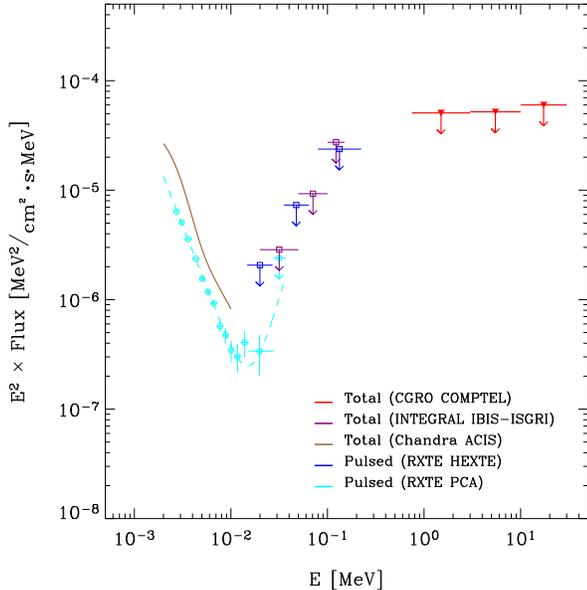}}
  \caption{\label{HE_SPECTRUM_1E}A $\nu F_{\nu}$ spectral representation of the total (2 keV - 30 MeV) and pulsed (2-30 keV) 
  high-energy emission from \axpe. The dashed aqua coloured line represents the double power-law fit to the PCA 2-30 keV pulsed 
  spectrum. This fit suggests the onset of a hard spectral tail in the pulsed spectrum near $\sim 15$ keV.}
\end{figure}

The ISGRI flux upper limits to the total emission from \axpe\ for energies above
20 keV \citep{hartog06} can be directly compared with the Chandra ACIS spectrum for the 2-10 keV range. 
The black-body plus power-law model fit to this spectrum \citep{patel01} is shown in Fig. \ref{HE_SPECTRUM_1E} 
(dark-orange solid line). The double-power-law spectral fit suggests that again a minimum
in luminosity is reached around 20 keV, where the extrapolation of the
Chandra model fit crosses the pulsed PCA spectrum. Given the hard pulsed component 
suggested in these PCA data up to $\sim 24$ keV, we estimate that the source
will show up at higher energies in an IBIS ISGRI mosaic totalling $\gtap 4$ Ms in effective 
on-axis exposure, what will be reached when adding in follow-up work already performed and/or
scheduled observations. To complete the high-energy picture, the COMPTEL 
0.75-30 MeV upper limits to the total emission \citep{hartog06} 
are also included in Fig. \ref{HE_SPECTRUM_1E}. 

Comparing in the overlapping energy range below 10 keV the PCA time-averaged pulsed fluxes with the 
Chandra total emission model, and assuming no time variability, indicates a rather high value of 
$\gtap 43\%$ for the pulsed fraction (pulsed/total emission), in agreement with earlier 
estimates \citep{patel01}.

\section{\axpee}

\axpee\ was discovered with {\it Einstein\/} at an angular separation of $\sim 40\arcmin$ from $\eta$ Car
\citep{seward82}. During an observation on July 13, 1979 a sinusoidally shaped signal with a high pulsed 
fraction of $68\pm 7 \%$  was detected at a period of $\sim 6.44$ s \citep{seward84,seward86}. The pulsations 
at a period of 6.4407(9) s were confirmed in EXOSAT ME observations \citep{smale85,seward86}, and the positional 
accuracy of \axpee\ was further tightened up down to a radius of about $10\arcsec$.

Ginga observations combined with earlier measurements revealed a steady increase in spin period 
at a mean rate of $\dot{\hbox{P}} = (4.64\pm 1.1) \times 10^{-4}$ s yr$^{-1}$ \citep{corbet90}, implying
a rotational energy loss much smaller than the lower limit on the X-ray luminosity. 

X-ray observations in the early nineties by ROSAT and ASCA \citep{mereghetti95,corbet97} indicated that the 
spin-down rate almost doubled after 1988, likely anticorrelated with the X-ray flux.

\begin{figure}[b]
 \centerline{\includegraphics[height=12cm,width=7cm,angle=0,bb=190 195 405 610]{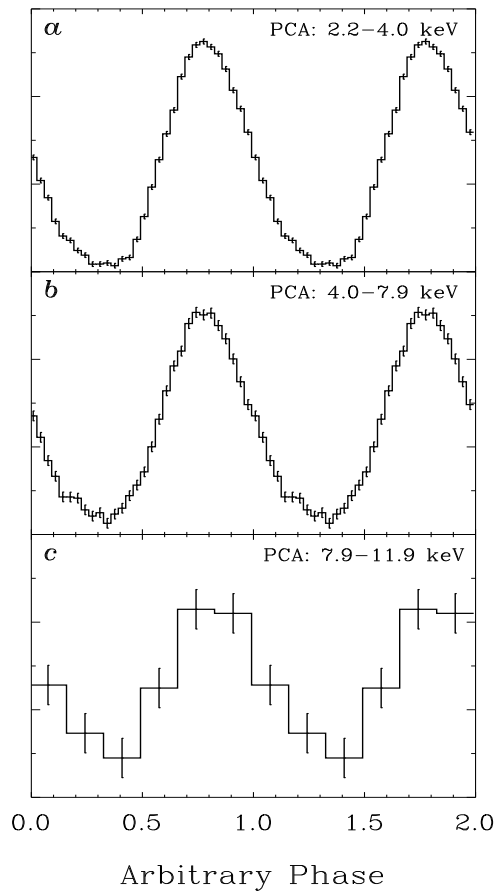}}
\caption{\label{pca_ee_stack}RXTE PCA pulse profiles of \axpee\ 
  for energies in the range 2.2-11.9 keV combining data collected between
  July 27, 1996 and Feb. 24, 2004 (see Table \ref{obs_table}). 
  Two cycles are shown for clarity.}
\end{figure}

BeppoSAX LECS / MECS data from long exposures in May 1997 indicated the necessity of a two component 
spectral model, a power-law plus black body, to describe properly the measured X-ray (0.5-10 keV) spectrum 
\citep{oosterbroek98}. This was confirmed by \citet{paul00} using ASCA data. In the 
meantime, detailed timing studies using RXTE monitoring observations performed in 1997-2000 \citep{kaspi01} showed 
that significant deviations from simple spin-down exist, making phase-coherent timing impossible over time stretches 
longer than a few months. Inspite of these rotational irregularities, neither pulse profile changes nor
large pulsed flux variations were found. \axpee\ exhibited three X-ray bursts, on 2001 Oct. 29, 2001 Nov. 14, and 2004 June 29,
all caught during RXTE monitoring observations \citep{gavriil02b,gavriil06}. 

Precise X-ray imaging of \axpee\ with Chandra \citep{wang02,israel02} provided a source position with sub-arcsecond 
accuracy. Within the $0\farcs7$ error circle only one faint near-IR source was found which clearly showed variability. 
A detailed variability study in the infrared and optical was undertaken by 
\citet{durant05}. They established the variable nature in the infrared and optical and found a possible anticorrelation 
with the X-ray pulsed flux \citep[see][for an extensive study of the X-ray pulsed flux variability and spin-down rate variations]{gavriil04b}.
At radio frequencies (21 cm) an expanding hydrogen shell centered on \axpee\ was detected \citep{gaensler05}, which 
can be interpreted as a wind bubble blown by a 30-40 M$_\sun$ star, likely the massive progenitor of \axpee.

Finally, the large effective area of XMM Newton, combined with accurate imaging, made it possible to perform in-depth 
(phase-resolved) spectral analyses not polluted by effects caused by the presence of strong nearby X-ray sources. 
XMM observed \axpee\ on three occasions and results are presented by \citet{tiengo02,mereghetti04} and \citet{tiengo05}. 
Comparing the three XMM observations revealed long-term flux and pulsed fraction variations in anti-correlation. The 
featureless spectral shape, however, remained more or less the same, and could be described by a combination of a power-law 
with photon index $\Gamma \sim 2.7-3.5$ and a blackbody with temperature $kT \sim 0.63$ keV. Phase-resolved spectroscopy 
clearly indicated that the spectrum is softer at pulse minimum and harder at pulse maximum with respect to the phase averaged spectrum.
 
In the next section we present for the first time the hard X-ray/soft $\gamma$-ray timing and spectral characteristics 
of \axpee\ above $\sim 8$ keV using a) all available public RXTE PCA and HEXTE data (see Table \ref{obs_table}), b) INTEGRAL 
IBIS ISGRI data (see Table \ref{obsint_table}) and c) CGRO COMPTEL 0.75-30 MeV data.

\subsection{\axpee\ timing characteristics}

In the timing analysis of the full set of PCA observations of \axpee\ listed in Table \ref{obs_table} according to the guidelines
given in Sect. \ref{rxte_timing}, we produced {\em time-averaged} pulse-phase distributions for energies between $\sim 2$ and 35 keV.
In the folding/correlation process (see Sect. \ref{rxte_timing}) we used the ephemerides given by \citet{kaspi01}. 
Significant pulsed emission has been detected up to $\sim 12$ keV. The PCA pulse profiles of \axpee\ are shown in Fig. \ref{pca_ee_stack} 
for the three following differential energy bands, 2.2-4.0, 4.0-7.9 and 7.9-11.9 keV. The detection of pulsed emission, with a non-uniformity 
significance of $6.8\sigma$, between $\sim$ 7.9 and 11.9 keV is reported for the first time.
The single peaked pulse profiles, with a slightly steeper rise than fall, do not show morphology changes as a function
of energy. HEXTE data do not show pulsed emission at energies above $\sim$ 15 keV. 

\begin{figure}[t]
\centerline{\includegraphics[height=8cm,width=8cm,angle=0]{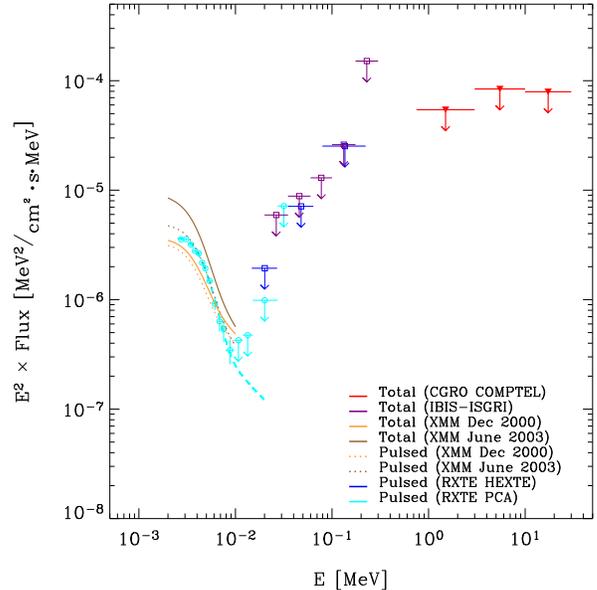}}
\caption{\label{HE_SPECTRUM_1E1048}A $\nu F_{\nu}$ representation of the spectrum
    of \axpee. Above $\sim 12$ keV the source is too weak to be detected so far. The IBIS ISGRI (purple) and CGRO
    COMPTEL (red) $2\sigma$ upper-limits are for the total emission. Below 10 keV we also added spectral information
    for the pulsed (dotted) and total (solid) emission from two XMM Newton observations performed at different
    epochs \citep{tiengo05}, illustrating the variable nature both in normalization and shape of both components.}
\end{figure}

\subsection{\axpee\ spectral characteristics}

\subsubsection{RXTE PCA/HEXTE pulsed spectrum}

The derived pulsed excess counts have been converted to flux values assuming an absorbing column density of 
N$_{\hbox{\scriptsize H}}=1.0 \times 10^{22}$ cm$^{-2}$ \citep[see][]{tiengo05}. It turned out that a single power-law model 
for the PCA band did not give an acceptable fit. Therefore we used for this AXP a black body plus power-law input spectrum. The 
optimized model ($\chi^2_r=0.55$ for 16 - 4 d.o.f.; $kT = 0.717(4)$ keV and $\Gamma=2.93(7)$) is shown in Fig. \ref{HE_SPECTRUM_1E1048} 
together with the flux measurements, both aqua colored. It is interesting to note, that, while the photon index is in the expected range \citep{tiengo05}, the (time-averaged) black body temperature of the pulsed component is slightly higher than the values obtained 
from the other X-ray instruments. 
This is very likely caused by the lack of sensitivity of the PCA for energies less than $\sim 2.5$ keV, thus poorly constraining the 
black body model at energies below its maximum. 
For HEXTE we could only derive upper limits for the pulsed emission, also shown in Fig. \ref{HE_SPECTRUM_1E1048} as blue 
symbols.

\subsubsection{INTEGRAL IBIS ISGRI and CGRO COMPTEL total spectrum}

IBIS ISGRI data (Core program -- GPS or public) from observations executed between revolution 30 and 217, satisfying
our off-source pointing constraint of $14\fdg5$ have been used to obtain spectral information on the total emission from 
\axpee\ in the 20 -- 300 keV energy range. The effective on-axis exposure time is, however, small $\sim$ 250.8 ks, resulting
in rather high upper-limits given the non-detections in any of the chosen broad energy bands.
These $2\sigma$ upper-limits have been included in Fig.\ref{HE_SPECTRUM_1E1048} as purple symbols. Also, the analysis of full
mission COMPTEL data (4.8 Ms exposure in total) did not result in significant detections of \axpee.
The $2\sigma$ COMPTEL upper-limits are shown as red symbols in Fig.\ref{HE_SPECTRUM_1E1048}. It is clear from the full 
high-energy spectral picture presented in Fig. \ref{HE_SPECTRUM_1E1048} that current spectral information above $\sim 10$ keV
does not exclude the presence of a hard spectral component. Additional IBIS ISGRI data e.g. from a deep 2 Ms observation of the 
Carina region, performed between INTEGRAL revolutions 192 and 203, with \axpee\ always in the FCFOV, will be very usefull to 
constrain the spectral properties of \axpee\ further in a future study.  


\section{\axp}

\axp\ is the first AXP for which surprisingly non-thermal pulsed X-ray/soft $\gamma$-ray emission
was discovered \citep{kuiper04}; see this paper also for a summary of earlier 
observational results on \axp). The pulsed spectrum above 10 keV could be described,
fitting RXTE PCA and HEXTE data, by a power-law model with photon index 
0.94 $\pm$ 0.16 up to $\sim$ 150 keV.  The total spectrum of \axp\ presented in
\citet{kuiper04} was derived from HEXTE data using source ON/OFF rocking-mode 
observations. We realized later that the HEXTE flux values were contaminated
by a contribution from the Galactic diffuse X-ray emission (see discussion in section 3.2).
Better spectral information can be obtained with an imaging instrument like INTEGRAL IBIS ISGRI. 
Preliminary total flux measurements with ISGRI were so far only reported by \citet{molkov04}.
For the present work significantly more ISGRI data are available. Therefore, we present in this 
section a) a new spectrum of the total emission and b) performed now also a timing analysis for 
\axp\ using ISGRI data. 
Furthermore, c) we repeated the timing/spectral RXTE PCA analysis using additional data
compared to \cite{kuiper04} as well as improved event selection criteria 
(see Sect. \ref{rxte_timing}). Finally, d) also for this source we analyzed COMPTEL skymaps.
 
\subsection{\axp\ pulse profiles for RXTE PCA and INTEGRAL IBIS ISGRI}
\label{sect_pca_tm}

\begin{figure}[t]
\centerline{\includegraphics[height=11cm,width=6.5cm,angle=0,bb=165 165 420 640]{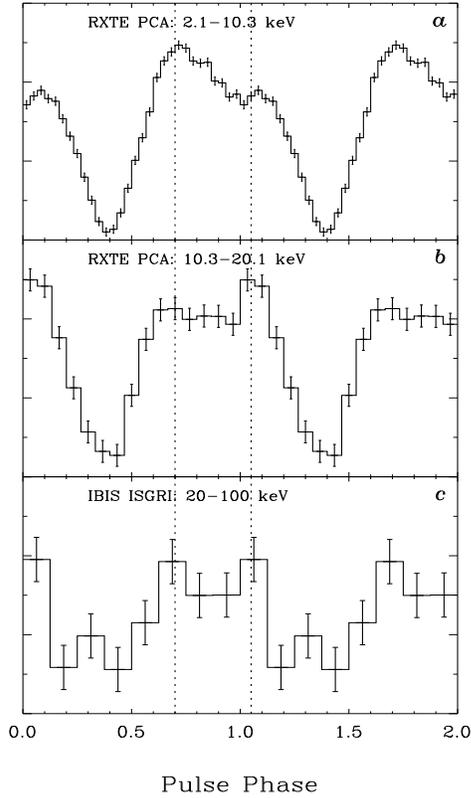}}
  \caption{\label{ISGRI_PCA_1E1841}RXTE PCA pulse profiles of \axp\ for the 2.1-10.3 and 10.3-20.1 keV energy ranges (panels a 
           and b, respectively), compared with the IBIS ISGRI 20-100 keV pulse profile (panel c).  
           The dotted lines at phases 0.7 and 1.05 serve as a guide to the eye for alignment comparison purposes. 
           The 20-100 keV IBIS ISGRI profile mimics the 10.3-20.1 keV RXTE PCA profile, suggesting a smooth prolongation
           above 20 keV of the 10.3-20.1 keV profile shape.}
\end{figure}

In our RXTE reanalysis for \axp, data from observation 80098 (spread over 28-3-2003 -- 17-2-2004; screened PCU-2 exposure 
$\sim 69.1$ ks) were added to the set used in \citet{kuiper04} (now, total screened PCU-2 exposure: $\sim 340.2$ ks). 
We used in the folding/correlation process (see Sect. \ref{rxte_timing}) the ephemerides given by \citet{gotthelf02} and derived in this work
(see Table \ref{eph_table}). 
The results are pulse profiles with improved statistics, confirming the variation of pulse shape with energy as presented by \citep{kuiper04}. As an example, Fig. \ref{ISGRI_PCA_1E1841} (panels a and b) shows RXTE PCA profiles for the 2.1-10.3 and 10.3-20.1 keV energy bands.

IBIS ISGRI data from observations with source off-axis angles less than $\le 14\fdg5$ and taken during INTEGRAL revolutions 49 - 123 have been
processed to yield pulse phase distributions for differential energy bands between 20 and 300 keV. The total screened effective on-axis exposure was $\sim 808.8$ ks. We derived a very accurate (RMS residual is 0.015 in phase; see Table \ref{eph_table} for the details) phase coherent ephemeris using RXTE PCA monitoring observations performed during March - December 2003. 
This has been applied in the folding process of the selected barycentered ISGRI events.
For the 20-100 keV energy range we obtained a $Z_2^2$ of $3.6\sigma$. This ISGRI pulse profile is also shown in Fig. \ref{ISGRI_PCA_1E1841} 
(panel c). Its shape mimics that of the PCA 10.3-20.1 keV profile suggesting a stable pulse shape in the hard X-ray window. The statistics 
are still low, however, but additional ISGRI data from already performed and future scheduled observations will allow more detailed studies 
of the pulse shape in this energy range.

\subsection{\axp\ spectral characteristics}

\begin{figure}[t]
\centerline{\includegraphics[height=8cm,width=8cm,angle=0]{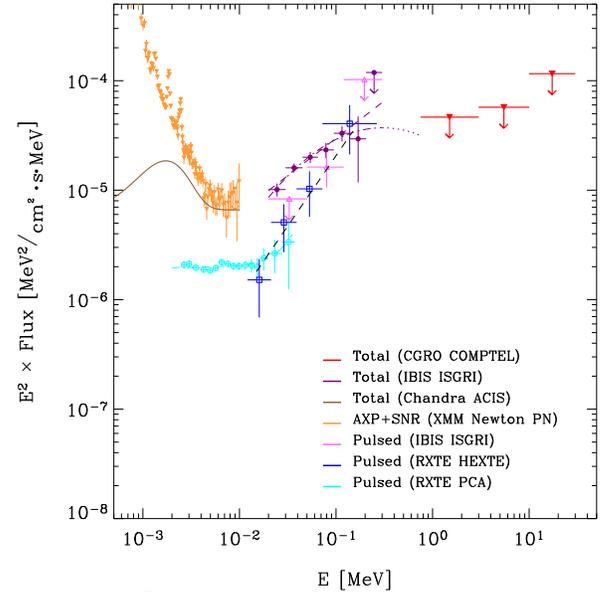}}
\caption{\label{HE_SPECTRUM_1E1841}A $\nu F_{\nu}$ spectral representation of the high-energy emission
  from \axp\ and SNR Kes 73. The aqua, blue and magenta data points/curves represent the (time averaged) pulsed emission component of
  the spectrum (2-300 keV) based on RXTE PCA/HEXTE and INTEGRAL IBIS ISGRI observations. The black dashed line
  shows the best power-law model fit to the PCA, HEXTE and ISGRI pulsed flux values for energies above $\sim 15$ keV.
  The purple data points/curves reflect the IBIS ISGRI total flux measurements and fits (power-law, dashed; curved power-law, 
  dashed-dotted and extrapolated to 750 keV). The red $2\sigma$ upper limits from CGRO COMPTEL require a spectral break, somewhere
  between 140 and 750 keV. For completeness, below 10 keV also the total flux from \axp\ plus SNR Kes 73 
  (yellow symbols; XMM Newton) and the total flux from \axp\ (dark orange solid line; Chandra ACIS) are shown \citep[see][for more details]{kuiper04}.}
\end{figure}

\subsubsection{RXTE and INTEGRAL pulsed emission}

Starting with the new RXTE PCA phase distributions, introduced above, the updated \axp\ pulsed spectrum, now 
with significant flux values up to $\sim 28$ keV, is shown in Fig. \ref{HE_SPECTRUM_1E1841} using aqua colored symbols. 
Also given are the earlier-reported pulsed-flux values from HEXTE \citep[dark blue data points; see][]{kuiper04}, which clearly revealed 
the hard pulsed component up to $\sim$ 150 keV.

Now we can also derive information on the pulsed spectrum from IBIS ISGRI. Phase distributions have been generated, followed 
by extraction of pulsed counts, for three logarithmically spaced energy bands (20-50, 50-120 and 120-300 keV), yielding
only for the 50-120 keV band a significant pulsed-flux measurement. This ISGRI pulsed-flux value and two upper limits, shown 
in Fig. \ref{HE_SPECTRUM_1E1841} as magenta symbols, are fully consistent with the hard spectrum measured by HEXTE.
We fitted the new PCA pulsed spectrum for energies above $\sim 15$ keV together with the HEXTE and INTEGRAL pulsed flux data points 
with a simple power-law model, yielding as best fit a photon index $\Gamma$ of $0.72 \pm 0.15$, slightly harder than derived by 
\citet{kuiper04}. This best fit model is indicated as a dashed black line in Fig. \ref{HE_SPECTRUM_1E1841}.

\subsubsection{INTEGRAL total emission}

In this work we derived the total emission spectrum of \axp\ in the 20-300 keV 
energy range from IBIS ISGRI mosaic maps generated for 7 differential energy 
bands over the 20 - 300 keV band, thus significantly improving upon the 
spectral information given by \citet{molkov04}. 
We used a combination of public and Core program observations, all with source 
off-pointing angles less than $14\fdg5$, and for which the
details\footnote{Data from Revs. 131,133-134 were ignored because these have 
been taken in a staring mode configuration giving rise to substantial systematical 
effects in the image deconvolution procedures.} are listed in Table 
\ref{obsint_table}. The total screened effective on-axis exposure on \axp\ in
this combination was $\sim 1,111$ Ms. Significant emission has been detected up 
to $\sim 140$ keV (95-140 keV; $5.7\pm 0.9$ mCrab). 
The derived flux values are shown in Fig. \ref{HE_SPECTRUM_1E1841} as purple 
colored symbols. Over this 20-300 keV band we can {\em not} discriminate
between a power-law or power-law model with energy dependent index to be the best 
model describing the measurements.
However, including in Fig. \ref{HE_SPECTRUM_1E1841} the CGRO COMPTEL flux 
upper limits (red symbols) derived in this work, we notice that the
best fitting power-law model with photon index $\Gamma= 1.32 \pm 0.11$ does not extend 
into the MeV domain. The spectrum must break somewhere between 140 and 750 keV. 
The curved power-law model is consistent with the COMPTEL $2\sigma$ upper limits.
Both models are shown in Fig. \ref{HE_SPECTRUM_1E1841} as purple dashed (power-law) and 
purple dashed-dotted (curved power-law) lines with the latter model extrapolated up to 750 keV.

\subsubsection{Discussion of spectra and pulsed fractions of \axp}

The new and better total spectrum of \axp, now derived using ISGRI data, is indeed
lower, $\sim$ 40 \% between 20 and 100 keV, than published earlier using HEXTE data 
\citep{kuiper04}, and is also slightly harder. It connects better to the total spectra 
measured at different epochs by XMM Newton and Chandra ACIS below 10 keV, shown in 
Fig. \ref{HE_SPECTRUM_1E1841} and adopted from \citet[][and citations therein]{kuiper04}.   

For the pulsed spectrum, the power-law fit to the new PCA spectrum above $\sim 15$ keV combined with
the new ISGRI measurements and the earlier published HEXTE flux values, confirms the drastic up turn/hardening 
of the pulsed spectrum. 
Also, the extrapolation of this very hard spectrum has to remain under our COMPTEL 
upper limits, requiring an even more drastic bend/break compared to what is 
required for the total spectrum.

The compilation of Chandra, XMM-Newton, RXTE and INTEGRAL spectra in Fig. \ref{HE_SPECTRUM_1E1841}, taken at 
very different epochs over many years, suggests that the hard X-ray emission of \axp\ is stable, i.e. this 
magnetar is most of the time in the same state of activity for energies above $\sim$ 10 keV. 
The pulsed fraction is confirmed to be $\sim$ 25\% at 20 keV and $\sim$ 100\% for energies beyond $\sim 100$ keV.


\section{Summary}
     
Exploiting the availability of archival data from RXTE monitoring observations
and new, deep INTEGRAL exposures of AXPs, we were able to show that AXPs
exhibit exceptionally hard spectra for energies above 10 keV. Of the sample of five AXPs 
studied in this work, three (the brightest at energies below 10 keV: \axp, \axprxs, \axpu) 
are shown to emit up to energies of $\sim$ 150 keV.
Of the two weaker sources, for one (\axpe) an upturn of pulsed emission is suggested 
up to $\sim$ 25 keV, and the other (\axpee) is too weak to detect a possible similar upturn 
in the hard-X-ray range with presently available exposures. 
We regard this sufficient evidence to conclude that a persistent non-thermal component 
at energies above 10 keV is a common property of AXPs.
 
\begin{table*}[t]
\caption{Pulsar characteristics and high-energy spectral properties (pulsed/total) of the AXPs studied in this work.\label{summary_table}}
{\scriptsize
\begin{center}
\begin{tabular}{|l|ccccc|cccc|cc|}
\hline
AXP & $P$ & $\dot{P}$ & $B_{s}$ & $L_{sd}^{32}$ & $d$ & $L_{1-10}^{32,p}$ & $\Gamma_l^p$ & $L_{10-100}^{32,p}$ & $\Gamma_h^p$ & $L_{10-100}^{32,T}$ & $\Gamma_h^T$  \\
          & s   & $10^{-11}$ s/s & $10^{14}$ G &  erg/s   & kpc & erg/s    &   & erg/s  &    & erg/s         &   \\
\hline
\axp          & 11.78 & 4.44  &  7.32  & 10.73  & 6.7 &  391  & $1.98(2)$      &     1312       &   $0.72(15)$   &   2975    &  $1.32(11)$     \\
\axprxs       & 11.00 & 1.92  &  4.64  &  5.68  &  5  &  640  & $2.60(1)$      &     719        &   $1.01(12)$   &   869     &  $1.44(45)$     \\
\axpu         &  8.69 & 0.20  &  1.34  &  1.21  &  3  &  347  & $4.09(2)$      &     686$^{\dagger}$ &   $-0.80(9)^{\dagger}$   &   638     &  $1.05(11)$     \\
\axpe         &  6.98 & 0.048 &  0.59  &  0.56  &  3  &  485  & $4.26(1)$      &      188$^{\dagger}$       &   $-1.02(19)^{\dagger}$  & $\ldots$  &  $\ldots$       \\
\axpee        &  6.45 & 2.31  &  3.90  & 33.90  & 2.7 &   76$^{\ast}$  & $^{\ast}$      &   $\ldots$     &   $\ldots$     & $\ldots$  &  $\ldots$       \\
\hline
\multicolumn{12}{l}{$^{\dagger}$ Based on double power-law fit to PCA data ($\sim$ 2-30 keV); 30-100 keV luminosity extrapolated from this model} \\
\multicolumn{12}{l}{$^{\ast}$     The 1-10 keV luminosity has been derived from a black body plus power-law model fit} \\
\multicolumn{12}{l}{$L^{32}$ means that the luminosity is expressed in units $10^{32}$ erg/s} \\
\end{tabular}
\end{center}}
\end{table*}

In Table \ref{summary_table} we summarize for the five AXPs the spectral parameters presented in this work,
together with estimates for their distances and pulsar characteristics (period, period derivative, deduced 
surface magnetic field strength at the pole and spin-down power).
Most interesting and new are the results on the total and pulsed spectra, and luminosities above 10 keV.
As noted above, the exceptionally hard pulsed spectra seem to extend up to at least 150 keV, therefore we have
given for all four AXPs for which the spectral up turn around 10 keV has been measured, the luminosity of the 
pulsed component integrating the best-fit power-law model over the full decade in energy from 10 to 100 keV. 
For the luminosity of the pulsed component between 1 and 10 keV, we list the power in the power-law component 
fitted to the pulsed spectra before the break, except for \axpee, where the power has been derived from a 
black body plus power-law model.
Thus for 4 AXPs we ignore the presence of a black-body component which is known to be present at lower energies as well.  
However, for \axprxs, \axpu, \axpe\ and \axp\ we could fit the pulsed PCA spectra down to about 2 keV well with 
a single power-law model. 

The total spectra above 20 keV appear also to be hard, but somewhat softer than the pulsed spectra. 

For \axp\ and \axpu\ the total emission becomes consistent with 100\% pulsed around $\sim 100$ keV, starting from a pulsed 
fraction at 10 keV of $\sim$ 25\% and $\sim$ 10\% for \axp\ and \axpu, respectively.
The exception is \axprxs, for which the hard X-ray emission above 10 keV is consistent with being 100\% pulsed. 
Note in this context, however, the strong intensity variability reported for this AXP for energies below 10 keV, which 
might be coupled to variations in pulsed fraction.

Table \ref{summary_table} immediately shows that the luminosities of the hard X-ray components (pulsed and total) 
exceed the available total spin-down powers by a few orders of magnitude, a conclusion drawn earlier for the total luminosities (black body 
plus power-law components) in the 1 -- 10 keV band. 

Our compilations of total and pulsed flux values for energies above 10 keV for the three
strongest AXPs, using data from different instruments and collected at different epochs,
suggest that the high-energy component is stable, contrary to reports for the emissions
below 10 keV.  However, addional observations are required (and will become available with INTEGRAL) 
to constrain this further.

The time-averaged light curves for the different AXPs suggest that the pulse shapes for energies above 10 keV exhibit 
less structure, and vary less with energy than seen below 10 keV. However, the statistics are still insufficient to draw 
stringent conclusions. Phase-resolved spectroscopy on time-averaged profiles will be performed when more INTEGRAL data 
can be added.

\section{Discussion}

The discussion of AXPs in the context of magnetically powered rotating neutron stars,
was till recently focussed on understanding the observational characteristics of the soft-spectrum 
component for energies below 10 keV. Indeed, evidence for hard X-ray or soft gamma-ray emission from 
AXPs was lacking, in contrast to the very luminous outbursts at soft gamma-ray energies characterizing SGRs. 
Our discovery of pulsed emission from AXPs up to at least $\sim$ 150 keV adds a completely new 
non-thermal component requiring a steady mechanism for accelerating particles in magnetospheres of magnetars. 
Also this new quiescent emission component  is far too luminous to be powered by rotational energy loss, as is evident 
in Table \ref{summary_table}. Recently, INTEGRAL also found for the first time quiescent hard X-ray emission from a SGR 
\citep[SGR 1806-20;][]{molkov05,mereghetti05}. As for AXPs, its spectrum above 10 keV could be fitted with a hard 
power-law model, photon index 1.6 $\pm$ 0.1, an index comparable to, but somewhat softer than that measured for the 
total AXP spectra in that energy window, but significantly softer than the hard spectra reported in this work 
for the quiescent pulsed spectra of AXPs. In addition, the quiescent SGR spectrum does not exhibit a black body component 
at energies below 10 keV. This underlines again, that AXPs and SGRs are (very) different manifestations of the magnetar scenario.

In Table \ref{summary_table}, we listed the high-energy characteristics for five AXPs studied in this work, out of a total of 
six persistent AXPs known to date. Although this sample is still very small, we verified whether there are already any apparent 
correlations between the measured parameters. There is a hint for a correlation between the magnetic field strength $B_s$
and the luminosity $L_{10-100}^p$ of the pulsed emission of the hard tail (10-100 keV) for four AXPs, as well as the total emission 
of the hard tail (now for three AXPs). However, \axpee\ has a $B_s$ three times higher than \axpu, and is located at a similar 
distance, but no hard X-ray emission has been detected, and its luminosity at $\sim$ 10 keV is about three times lower than that 
of \axpu. It is interesting to note that the spin down luminosity of \axpee\ is the highest of this sample, mainly due to its large $\dot P$, 
and reaches almost 50\% of the luminosity of the pulsed emission between 1 and 10 keV, a factor 10 - 100 higher than for the other four AXPs.

\begin{figure}[t]
\centerline{\includegraphics[height=8cm,width=8cm,angle=0]{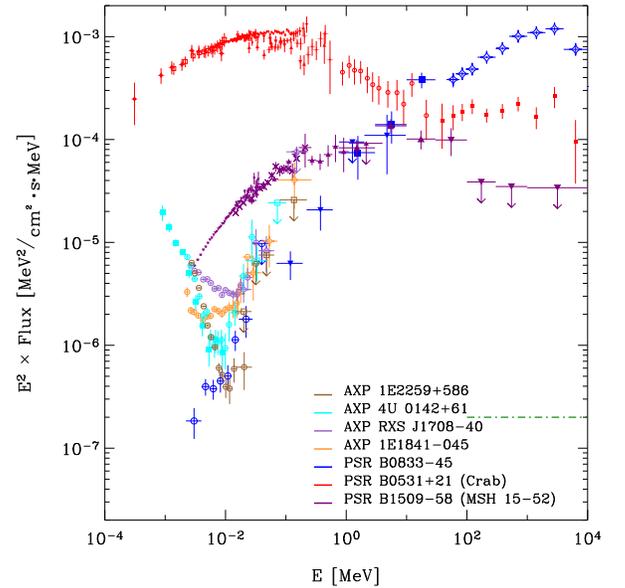}}
\caption{\label{HE_SPECTRUM_AXP_HE} A $\nu F_{\nu}$ spectral representation of the pulsed emission for the 4 AXPs showing 
  hard spectral tails. For comparison purposes, also the pulsed high-energy ($\sim$ 1 keV - 10 GeV) spectra of the young Crab (PSR B0531+21)
  and PSR B1509-58 pulsars, and the middle-aged Vela pulsar (PSR B0833-045) are shown. Furthermore, the $3\sigma$ GLAST sensitivity 
  (green dashed-dotted line) assuming an $E^{-2}$ source for an all-sky survey duration of one year ($>100$ MeV) is shown.}
\end{figure}

Surprisingly, the high-energy AXP spectra are very similar to those of "middle-aged" radio pulsars in the $\sim$ 10--300 keV 
energy band. This can be seen in Fig. \ref{HE_SPECTRUM_AXP_HE} in which the AXP spectra are shown together with the spectra 
of two young radio pulsars, PSR B0531+21 (Crab) and PSR B1509-58, and the middle-aged Vela pulsar (PSR B0833-45). These are 
the only radio pulsars which have been detected from soft X-rays up to the gamma-ray domain. The other middle-aged high-energy 
gamma-ray pulsars detected by CGRO EGRET \citep[see e.g. the review by][]{thompson97} are detected at high-energy gamma-rays 
($>$100 MeV), and soft X-rays, but not in the intermediate hard X-ray range, in which they are weaker than the Vela pulsar. 
The overall high-energy spectral shape of these middle-aged pulsars is expected to be similar to that of Vela. Like the AXPs, 
the Vela pulsar has a (pulsed) thermal black-body component (only visible below 2 keV, and not shown in Fig.\ref{HE_SPECTRUM_AXP_HE}), 
and an extrapolation of its power-law shape spectrum above 10 keV to the IR, optical and radio domains is approximately in agreement
with the measured flux values at these longer wavelengths. This is rather similar to the case of e.g. AXP \axpu, 
with the difference that none of the AXPs have so far been seen in the radio, and the X-ray and IR emissions
exhibit a time variable behaviour, not seen for Vela.  This similarity of the multiwavelengths spectra could suggest 
some similarity in production processes in the two different scenarios of rotation-powered pulsars and magnetars
starting from very different sources of power.

A first attempt, before our discovery of hard X-ray emission from AXPs, to model possible production of high-energy (X-ray and gamma-ray) 
emission in the magnetospheres of AXPs, applying a scenario proposed for high-energy production in the magnetospheres of radio pulsars, was made by
\citet{cheng01}. They modeled the production of high-energy (above 100 MeV) gamma-radiation in outer magnetospheric gaps of AXPs. 
They argued that due to the strong field of a magnetar, the gamma-ray emission rooted at the polar caps will be quenched. 
However, far away from the pulsar surface, i.e. in outer vacuum gaps, gamma radiation could be emitted because the local field will 
drop below the critical quantum limit. They predict that the NASA Gamma-ray Large Area Space Telescope GLAST, due for launch in 2007, 
might be able to detect this very-high-energy emission from AXPs (see Fig. \ref{HE_SPECTRUM_AXP_HE} for the $3\sigma$ GLAST sensitivity 
assuming an $E^{-2}$ source spectrum for an all-sky survey duration of one year). However, their model calculations do not reproduce 
the hard spectra and high X-ray luminosities now discovered above 10 keV.

The publication of the discovery of a hard spectral tail of \axp\ by \citet{kuiper04} 
stimulated \citet{thompson05} to reconsider high-energy emission from magnetars due to an 
ultra strong magnetic field. A gradual release of energy in the stellar magnetosphere is expected 
if it is twisted and a strong electric current is induced on the closed field lines. They 
considered two mechanisms of gamma-ray emission:
 
(1) A thin surface layer of the star is heated by the downward beam of current carrying
charges, which excite Langmuir turbulance in the layer. Thus, a temperature kT of $\sim$ 
100 keV can be reached, emitting bremsstrahlung photons up to this characteristic temperature. 
Interestingly, three of the AXPs, \axprxs, \axpu\ and \axp, were detected up to these energies of $\sim$ 150 keV, 
the shapes of their total and/or pulsed spectra being consistent with this mechanism with peak 
luminosities around the predicted energy. However, the $100\%$ pulsed fraction around 20 keV for \axprxs\ is difficult to 
reconcile with the surface emission, given the effects of general relativistic light bending.
Future observations have to reveal whether the cut off in the spectra is also consistent with this interpretation.    

(2) Soft $\gamma$-rays are produced at a distance of $\sim$ 100 km
from the star surface in the magnetosphere, where the electron cyclotron energy
is in the keV range. A large electric field develops in this region in response to
the outward drag force felt by the current-carrying electrons from the flux of keV
photons leaving the star. Thompson \& Beloborodov show that a seed photon
injected in this region undergoes a runaway acceleration and upscatters keV 
photons above the threshold for pair creation. The resulting synchrotron spectrum can 
reach a peak at $\sim$ 1 MeV.  Our hard-X-ray spectra together with 
the upper limits from COMPTEL in the MeV domain,  allow spectra with
maximum luminosities close to the MeV range.
But we should note that the flux values reported around $\sim$ 100 keV are already
at about the same level as the 2-$\sigma$ upper limits for the 0.75--3 MeV interval,
suggesting a different shape, thus challenging this interpretation.

An alternative quantum-electrodynamics model for the bursts and the quiescent 
non-thermal emission from AXPs as well as SGRs was proposed by \citet{heyl05a,heyl05b}.  
Their model is based on fast-mode breakdown, in which wave energy is dissipated into 
electron-positron pairs when the scale of these discontinuities becomes comparable to an 
electron Compton wavelength. They showed that under appropriate conditions, an electron-positron
fireball would result, producing primarily the X-ray and soft-gamma-ray bursts.
They also investigated \citep{heyl05b} the properties of non-thermal emission in the absence 
of a fire ball. This quiescent, non-thermal emission associated with fast-mode breakdown 
may account for the observed non-thermal emission presented in this work for AXPs, as well as for the quiescent
emission reported for SGR 1806-20 \citep{molkov05,mereghetti05}.
Indeed, they succeeded in fitting ISGRI AXP spectra as well as unabsorbed
optical data for \axpu\ from \citet{hulleman00}, and the SGR 1806-20
non-thermal spectrum above 10 keV. Interestingly, they predict that the
emission should extend beyond the observed INTEGRAL spectra without a break below 
1 MeV. However, the combination of ISGRI spectra and COMPTEL upper limits presented 
in this work, seem to contradict this claim. They further state, that if the magnetars
have significant optical excesses, such as \axpu, then the quiescent emission
from most of the AXPs discussed in this work and of SGR 1806-20 should be detectable
with GLAST (see again Fig. \ref{HE_SPECTRUM_AXP_HE} for the GLAST sensitivity). 
Further INTEGRAL observations, and ultimately GLAST observations can support or reject these claims.

Over the last few years, two radio pulsars with periods and period time derivatives
similar to those found for AXPs, thus also with high-magnetic-field strengths, are 
found: PSR J1847-0130, which has the highest by far inferred surface dipolar magnetic 
field (B = $9.4 \times 10^{13}$ G) among all known radio pulsars \citep{mclaughlin03} and 
PSR J1718-3718 \citep{hobbs04,kaspimcl05}.
The X-ray luminosities (or upper-limits to these) in the 2-10 keV range derived for these 
two radio pulsars are much lower than those of the standard AXP group. Above 10 keV, detection of 
hard X-rays could connect these two different types of neutron stars.
PSR J1847-0130 and PSR J1718-3718 both happened to be located near the centers of our 
deep IBIS ISGRI exposures on \axp\ and \axprxs, respectively. Therefore, we searched above 20 keV
for hard X-ray signatures from these pulsars, but nothing was detected in any of the energy bands 
we have investigated. The $2\sigma$ upper limit on the flux in the 20-30 keV band for PSR J1847-0130 is 
0.4 mCrab\footnote{\axp\ is detected in the same image with a flux of $\sim 1.5\pm 0.2$ mCrab} 
and that for PSR J1718-3718 is also 0.4 mCrab, but now in the 20-35 keV band.
 
Therefore, why solitary rotating neutron stars with AXP-like timing parameters in some cases 
behave as `'dull" radio pulsars and in other cases as `'enigmatic" magnetars is still unclear. 
Different suggestions have been made \citep[see e.g.][]{mclaughlin03}, one of the possibilities being
that high-field pulsars and AXPs have similar dipole magnetic fields, but
AXPs have also quadrupole (or higher) components.
\citet{kaspimcl05} suggest a.o. the interesting possibility that
the high-field radio pulsars, may one day emit transient magnetar-like emission,
and conversely that the transient AXPs might be more likely to exhibit
radio pulsations. More theoretical work is required, but certainly also more
detailed observational results at all wavelengths to verify the model predictions.
Ongoing observations of AXPs with the wide-field-of-view INTEGRAL IBIS imager will allow us to
contribute with deeper studies in the hard X-ray/soft $\gamma$-ray range.


\acknowledgments

This research has made use of data obtained from the High Energy Astrophysics 
Science Archive Research Center (HEASARC), provided by NASA's Goddard Space Flight Center,
and of data obtained through the INTEGRAL Science Data Centre (ISDC), Versoix, Switzerland.
INTEGRAL is an ESA project with instruments and science data centre funded by ESA member 
states (especially the PI countries: Denmark, France, Germany, Italy, Switzerland, Spain), 
Czech Republic and Poland, and with the participation of Russia and the USA.
We have extensively used NASA's Astrophysics Data System (ADS).
We thank Anton Klumper and Jaap Schuurmans for the maintenance of the OSA software at SRON and
the INTEGRAL data import from the ISDC site. Nanda Rea is acknowledged for useful discussions
on AXPs in general. Finally, we appreciate the willingness of Jacco Vink and Maurizio Falanga to 
make their INTEGRAL open time data of the Cassiopeia region (Revs. 261-269) directly available for us.

\clearpage 

\end{document}